\pdfoutput=1
\documentclass[a4paper,fleqn,usenatbib,umoline]{mnras}
\usepackage{amsmath}  % Advanced maths commands
\usepackage{amssymb}  % Extra maths symbols
\usepackage{graphicx} % Including figure files
\usepackage{txfonts}
\usepackage{xspace}
\usepackage{ae,aecompl}
\usepackage[dvipsnames]{xcolor}
\usepackage[multidot]{grffile}
\usepackage[capitalise]{cleveref}
\crefname{eq:}{Eq.}{Eqs.}
\crefname{fig:}{Fig}{Figs}
\usepackage{subfig}
\usepackage{upgreek}
\usepackage[normalem]{ulem}
\usepackage{wasysym}
\usepackage{booktabs}
\usepackage{threeparttable}  
\usepackage{lipsum}
\usepackage{soul}  
\newcommand{\taba}{\citetalias{Tabatabaei2016}\xspace}

\newcommand{\be}{\begin{equation}}
\newcommand{\ee}{\end{equation}}

\newcommand{\kelvin}{\,{\mathrm{K}}\xspace}

\newcommand{\re}{R_{\star}\xspace}
\newcommand{\rh}{R_\mathrm{h}\xspace}
\newcommand{\vre}{v_{\mathrm{\re}}}
\newcommand{\vrot}{v_{\rm rot}}

\newcommand{\Omegare}{\Omega_{\re}\xspace}

\newcommand{\Mtotre}{M_{\mathrm{tot,\,\re}}\xspace}
\newcommand{\kms}{\mbox{km s$^{-1}$}}

\newcommand{\msun}{\,M$_{\odot}$\xspace}
\newcommand{\msunyr}{M$_{\odot}$\,yr$^{-1}$\xspace}
\newcommand{\bbar}{\overline{B}}
\newcommand{\jcirc}{j_\mathrm{circ}\xspace}

\newcommand{\bre}{\overline{B}_\mathrm{\re}\xspace}
\newcommand{\btot}{\overline{B}_\mathrm{tot}\xspace}
\newcommand{\btoto}{\overline{B}_\mathrm{tot,\,0}\xspace}
\newcommand{\btotth}{\overline{B}_\mathrm{tot,\,30}\xspace}
\newcommand{\bturb}{\overline{B}_\mathrm{turb}\xspace}
\newcommand{\bturbth}{\overline{B}_\mathrm{turb,\,30}\xspace}
\newcommand{\bord}{\overline{B}_{\rm ord}\xspace}
\newcommand{\bordth}{\overline{B}_{\rm ord,\,30}\xspace}
\newcommand{\bordo}{\overline{B}_{\rm{ord,\,0}}\xspace}

\newcommand{\bturbre}{\overline{B}_\mathrm{turb,\re}\xspace}

\newcommand{\bcgd}{\overline{B}_\mathrm{cg,\,d}\xspace}
\newcommand{\bcgre}{\overline{B}_\mathrm{cg,\,\re}\xspace}
\newcommand{\bd}{\overline{B}_\mathrm{d}\xspace}
\newcommand{\Tcg}{T_\mathrm{cg}\xspace}

\newcommand{\Mstar}{M_\star\xspace}

\newcommand{\Msre}{M_\mathrm{\star,\,\re}\xspace}
\newcommand{\Mgre}{M_\mathrm{gas,\,\re}\xspace}

\newcommand{\Riso}{R_\mathrm{iso}\xspace}
\newcommand{\Spearman}{r_\mathrm{s}\xspace}
\newcommand{\Pearson}{r_\mathrm{p}\xspace}
\newcommand{\Spearmansf}{r_\mathrm{s,\,sf}\xspace}
\newcommand{\Pearsonsf}{r_\mathrm{p,\,sf}\xspace}
\newcommand{\asf}{a_\mathrm{sf}\xspace}
\newcommand{\sfrre}{\rm{SFR_{R_\star}}\xspace}

\newcommand{\Mbh}{M_{\mathrm{BH}}\xspace}

\newcommand{\fgas}{f_{\mathrm{gas}}\xspace}
\newcommand{\fbar}{f_{\mathrm{baryon}}\xspace}

\newcommand{\ijk}{{\mathrm{i,\,j,\,k}}\xspace}

\newcommand{\SI}{S_{\rm I}\xspace}
\newcommand{\SPI}{S_{\rm PI}\xspace}
\newcommand{\Ncr}{N_{\rm cr}\xspace}

\defcitealias{Tabatabaei2016}{T16}

\title[Reconciling magnetic fields with dynamical mass]{Reconciling the magnetic field in central disc galaxies with the dynamical mass using the cosmological simulations
}

\author[Hosseinirad et al.]{
Mohammad Hosseinirad,$^{1}$\thanks{E-mail: m.rad@ipm.ir (MH)}
Fatemeh Tabatabaei,$^{1}$
Mojtaba Raouf$^{2}$
and 
\newauthor
Mahmood Roshan$^{3,1}$
\\
 $^{1}$School of Astronomy, Institute for Research in Fundamental Sciences (IPM), PO Box 19395-5531, Tehran, Iran\\
 $^{2}$Leiden Observatory, Leiden University, P.O. Box 9513, 2300 RA Leiden, Netherlands\\
 $^{3}$Department of Physics, School of Sciences, Ferdowsi University of Mashhad, Mashhad, PO Box 91775-1436, Iran
}
\date{Accepted . Received ; in original form }

\pubyear{2021}

\begin{document}
\label{firstpage}
\pagerange{\pageref{firstpage}--\pageref{lastpage}}
\maketitle
\begin{abstract}
The Universe is pervaded by magnetic fields in different scales, although for simplicity, they are ignored in most cosmological simulations. In this paper, we use the TNG50, which is a large cosmological galaxy formation simulation that incorporates magnetic fields with an unprecedented resolution. We study the correlation of the magnetic field with various galaxy properties such as the total, stellar and gaseous mass, circular velocity, size and star formation rate. We find a linear correlation between the average magnetic field pervading the disc of galaxies in relative isolation and their circular velocities. In addition we observed that in this sample the average magnetic field in the disc is correlated with the total mass as $\bbar\sim\Mtotre^{0.2}$. We also find that the massive galaxies with active wind-driven black hole feedback, do not follow this trend, as their magnetic field is substantially affected by this feedback mode in the TNG50 simulation. We show that the correlation of the magnetic field with the star formation rate is a little weaker than the circular velocity. Moreover, we compare the magnetic field components of the above sample with a compiled observational sample of non-cluster non-interacting nearby galaxies. Similar to the observation, we find a coupling between the ordered magnetic field and the circular velocity of the flat part of the rotation curve in the simulation, although contrary to the observation, the ordered component is dominant in the simulation.
\end{abstract}


%%%%%%%%%%%%%%%%%%%%%%%%%%%%%%%%%%%%%%%%%%%%%%%%%%
\section{Introduction}
Magnetic fields are observed on all scales, from planets to stars to galaxies, and even in clusters of galaxies. It has been shown that magnetic fields constitute an energetic component of the interstellar medium (ISM) in galaxies and hence can possibly affect the formation and evolution of galactic structures such as spiral arms \citep[e.g.][]{Dobbs2008,Kotarba2009,Khoperskov2018} and galaxy centres \citep{Tabatabaei2018}. They can also play a role in driving winds and outflows \citep[e.g.][]{Steinwandel2020,Taba_2022}.

The origin of the magnetic fields is still unknown and a number of scenarios have been proposed for its creation \citep{Widrow2002,Kandus2011,Subramanian2016}. In some scenarios feedback plays an important role, in which magnetic fields can be generated by galactic winds \citep{Volk2000,Donnert2008a} or Active Galactic Nuclei (AGN) ejecta \citep{Furlanetto2001} at low redshifts, or seeded at high redshifts and then amplified by accretion-driven shear or adiabatic compression \citep{Dolag2005}. These seed fields could have astrophysical nature like outflow from dwarf starburst galaxies \citep{Kronberg1999} or be created in the early Universe \citep[see][for a review]{Durrer2013}. In another scenario, the merger shocks induced by the hierarchical structure formation could generate magnetic fields in a so-called Biermann-battery process \citep{Kulsrud1999,Ryu1998}. Nevertheless, the problem with the current available scenarios is the 25 orders of magnitude difference they predict for the primordial seed fields ($\sim 10^{-28}-10^{-3}\ \rm \mu G$) \citep{Dolag2006a,Vazza2021a}. It is clear that the tiny initial magnetic field must have been amplified afterwards to the present-day values through a variety of mechanisms, such as turbulent small-scale dynamos \citep{Kulsrud1999,Arshakian2009,Schleicher2013a}, large-scale galactic dynamo, or shear flows at different scales \citep{Dolag1999}.

Magnetic fields in galaxies have been the subject of many numerical studies. It involves modeling a small portion of galactic discs using a local shearing box in order to achieve the highest resolution and to resolve sophisticated physics \cite[e.g.][]{Kim2017b,Kim2020} , or simulating isolated disk galaxies in order to provide a comprehensive overview of galaxy evolution, including disc formation and fragmentation \citep{Peng2009}. Additionally, it has been shown the effects of seed fields, divergence cleaning schemes, and star formation \citep{Pakmor2013}, as well as the role of strong supernovae and radiation feedback on amplification of magnetic field \citep{Rieder2016}. Moreover, there have been studies that have demonstrated the injection of magnetic field by supernova explosion into the galaxy inside the ISM \citep{Butsky2017}, and the influence which spiral structures initially have on the enhancement of the magnetic field \citep{Khoperskov2018}. The \citet{Ntormousi2018} study also demonstrated the different configurations of the initial magnetic field with the help of sink particles, and \cite{Wibking2021} study presented the mapping of the vertical structure and topology of a magnetic field.

Using the more realistic zoom-in technique, galactic magnetic fields have been examined for the first time by \citet{Pakmor2014} in the full cosmological context, and it was found that the seed field grows exponentially until z=4 when it reaches saturation. In a subsequent paper, \citet{Pakmor2017} found that the linear amplification caused by the differential rotation, usually saturates at $z<0.5$. Moreover, \citet{Rieder2017} concluded that turbulent magnetic field is likely the dominant component in feedback-dominated galaxies at high-redshifts. This is also reported by \citet{Martin-Alvarez2018} who additionally showed that a strong primordial seed field retards the star formation while reducing the rotational support of galaxies and their sizes \citep{Martin-Alvarez2020} and potentially could pervade the intergalactic medium without mixing with galactic sources \citep[e.g.][]{Martin-Alvarez2021}. An interesting work by \citet{VandeVoort2020} indicates that magnetic fields can reduce the speed at which gas flows into the circumgalactic medium (CGM), resulting in a more metal-poor and more massive disc.
The zoom-in method has been also exploited for simulation of galaxy mergers, e.g., in a paper by \citet{Whittingham2021} where they demonstrated that magnetic fields can considerably augment the spiral structures and help to develop a more extended disc.

In the last approach, the rise of petascale supercomputers as well as advancements in algorithms, has made it possible to add the magnetic field to the large-scale cosmological structure formation simulations. This includes studies that are focusing on e.g., cosmic filaments and galaxy clusters \citep{Vazza2014,Marinacci2015,Aramburo-Garcia2021,Mtchedlidze2022}, intracluster medium \citep{Dolag2016}, galaxy population \citep{Marinacci2016}, or more recently galaxy formation during the epoch of re-ionization \citep{Katz2021}.The state-of-the-art IllustrisTNG project \citep{Marinacci2018a} is the first and currently the only one that follows galaxy formation in a large cosmological box by solving the magnetohydrodynamic equations from the very beginning of the Universe to the current time.

Investigating correlations between the magnetic field and other galaxy properties can shed light on its origin and impact on galaxy evolution. In this study, we use the highest resolution of the IllustrisTNG runs that can provide the possibility to decipher scaling relations between the magnetic field and main galaxy properties such as the total mass, rotation velocity and star formation rate, in large galaxy samples of different types. It is our intent to investigate these scaling relations in a sample of central (aiming to minimize environmental effects) disc galaxies with a view to comparing the results with similar galaxies observed.
More specifically, radio polarization observations indicate a tight correlation between the large-scale magnetic field and rotation speed of galaxies (\citet{Tabatabaei2016}, hereafter \taba). This empirical correlation is linked to a correlation with dynamical mass which is likely caused by a coupling between gas and magnetic field and flux freezing due to shear and density waves. An absence of a tight correlation with angular velocity is in favor of this explanation and a quenched dynamo amplification. Observations are, however, limited by galaxy sample size and polarization detection. Hence, it would be insightful to investigate the correlation found by \taba using simulations. For this purpose, we select galaxies with relative isolation from each other in the simulation box to mimic the nearby non-cluster, non-interacting galaxies studied by \taba. 

The structure of the paper is as follows: We first describe the IllustrisTNG simulation methods and its magnetic field implementation in \cref{sec:TNG,sec:B_in_TNG}.
Then, we explain our galaxy sample selection in \cref{sec:galaxy_sample}.
A temperature threshold is introduced for the definition of cold gas phase in the simulation (\cref{sec:cold_gas_def}) followed by designation of star-forming and quenched galaxies (\cref{sec:sf_or_q}).
After that, in \cref{sec:B_corr_TNG}, we try to search through this simulation for the existence of correlation between the magnetic field and dynamical mass as well as with other galaxy properties. In \cref{sec:BH,sec:different_T}, we specially shed light on the effect of temperature threshold and AGN feedback on the coupling of magnetic field with galaxy properties. Providing a clear context for comparing the observations with the simulation, we discuss and compare simulation with real observational results in \cref{sec:compare_with_T16,sec:ord_to_turb_ratio}. Moreover, issues related to the magneto-hydrodynamics (MHD) numeric that should be considered when interpreting the simulation outcomes are discussed in \cref{sec:mhd_problems}.  Our results are then summarised in \cref{sec:summary}.
%%%%%%%%%%%%%%%%% BODY OF PAPER %%%%%%%%%%%%%%%%%%
\section{Methods}
\subsection{The IllustrisTNG simulation}\label{sec:TNG}
The IllustrisTNG\footnote{\url{https://www.tng-project.org}} (hereafter TNG) is a set of cosmological magnetohydrodynamical simulations of galaxy formation \citep{Pillepich2018b,Naiman2018,Marinacci2018a,Springel2018,Nelson2018a,Nelson2018b}. It has been run in three main periodic boxes of $\simeq$ 300, 100 and 50 Mpc (comoving) side lengths (TNG300, TNG100 and TNG50), each with different resolutions. To follow $\Lambda$CDM cosmology, the TNG simulations adopted \citep{PlanckCollaboration2016a} cosmological set of parameters, assuming $\Omega_{\Lambda,0} = 0.6911$, $\Omega_{m,0} = 0.3089$, $\Omega_{b,0} = 0.0486$, $\sigma_8 = 0.8159$, $n_s = 0.9667$ and $h = 0.6774$. In this work, we use its highest resolution run available in the set, namely TNG50-1 (hereafter TNG50) \citep{Pillepich2019,Nelson2019} in which the dark matter (DM) and stellar particle /\ mean baryonic gas cell mass are $4.5\times 10^5$ and $8.5\times 10^4$\msun, respectively. At $z=0$, the softening length of DM and star particles is $\sim 288$ pc while this is $\sim 72$ pc for gas at its minimum value.

The TNG simulations are performed using periodic boundary conditions in Newtonian cosmological framework. In this simulation, the ideal MHD equations are solved with a quasi-Lagrangian scheme using the \textsc{arepo} code \citep{Springel2010a}  which exploits finite volume method on a moving unstructured Voronoi tesselation of the computation domain. The Poisson's gravity equation is solved by employing a Tree-Particle-Mesh (Tree-PM) \citep{Xu1995,Bode2000,Bagla2002} which computes the contribution of short- and long-range forces by using its tree and particle-mesh algorithms, respectively. To do so, Voronoi gas cells are considered as particles at their centre of mass and all other matter components.
\begin{figure}
	\centering
	\includegraphics[scale=0.43]{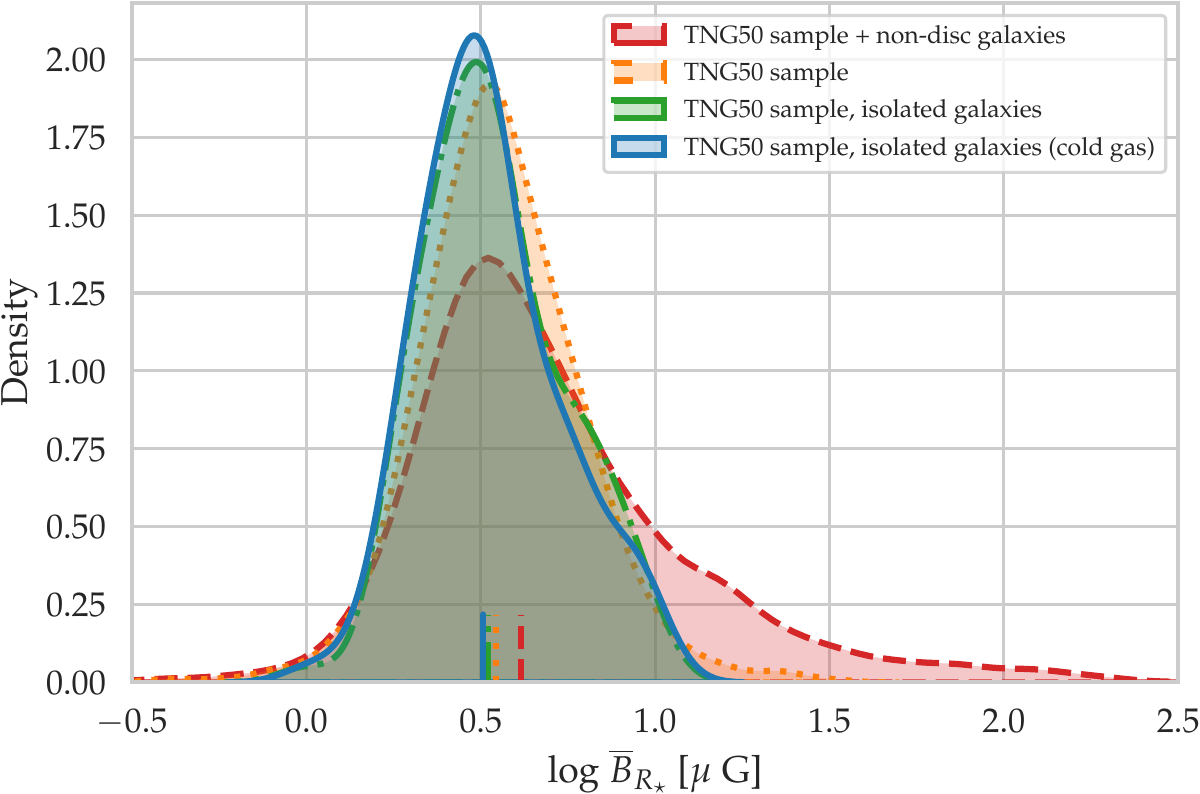}
	\caption{Comparison of the probability density function(PDF) of $\bre$ (the square root of the volume weighted value of $B^2$ in the effective radius $\re$) for the sample of central disc and non-disc galaxies in the TNG50 box, that have non-zero gas component or magnetic field (4547 galaxies; red dashed curve), the TNG50 sample (1697 galaxies; orange dotted curve), galaxies in the TNG50 sample which are isolated (103 galaxies; green dash-dotted curve) and the same galaxies whose  magnetic fields are calculated only in the cold gas cells (103 galaxies; blue solid curve). The small vertical lines show the corresponding median values.}
	\label{fig:kde_density_plot_B}
\end{figure}
\begin{figure*}
	\centering
	\includegraphics[scale=0.4]{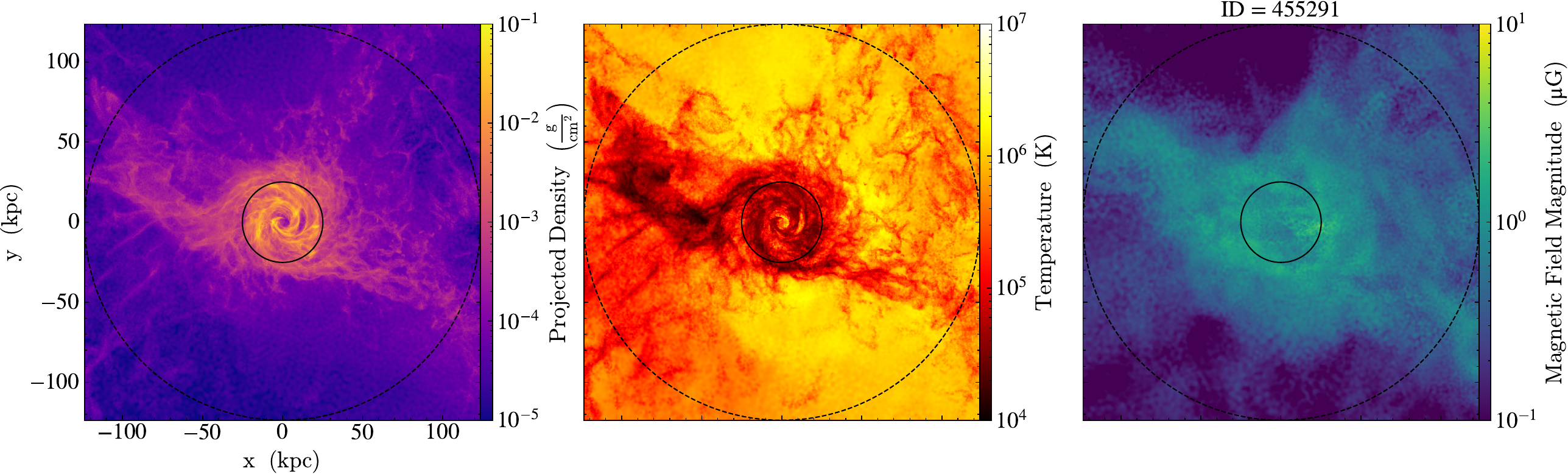}
	\includegraphics[scale=0.4]{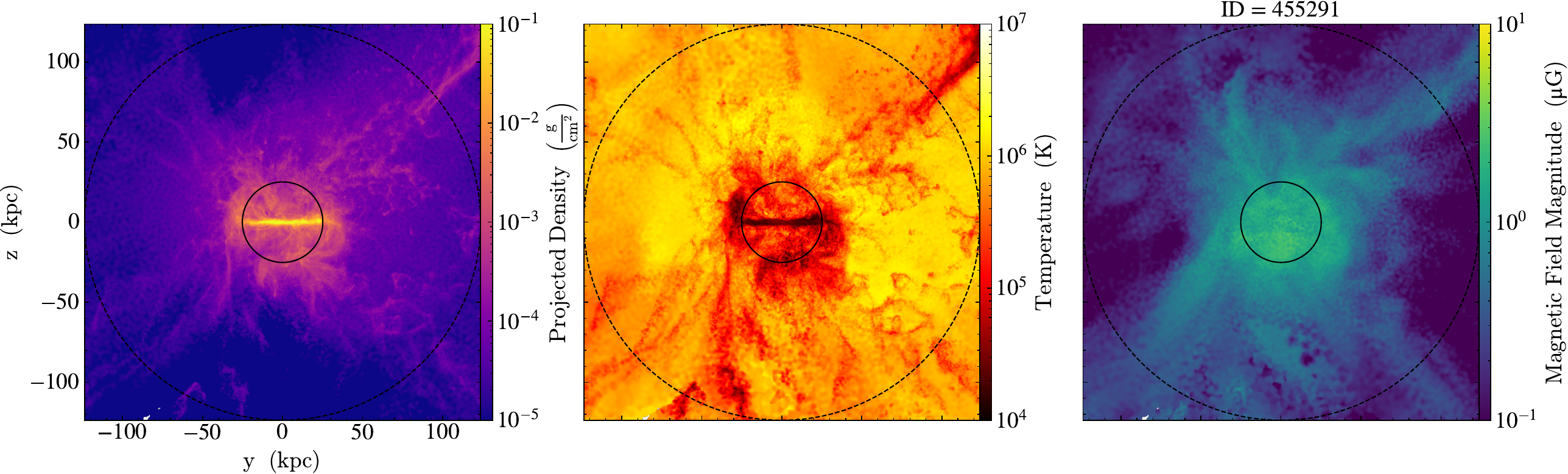}
	\includegraphics[scale=0.4]{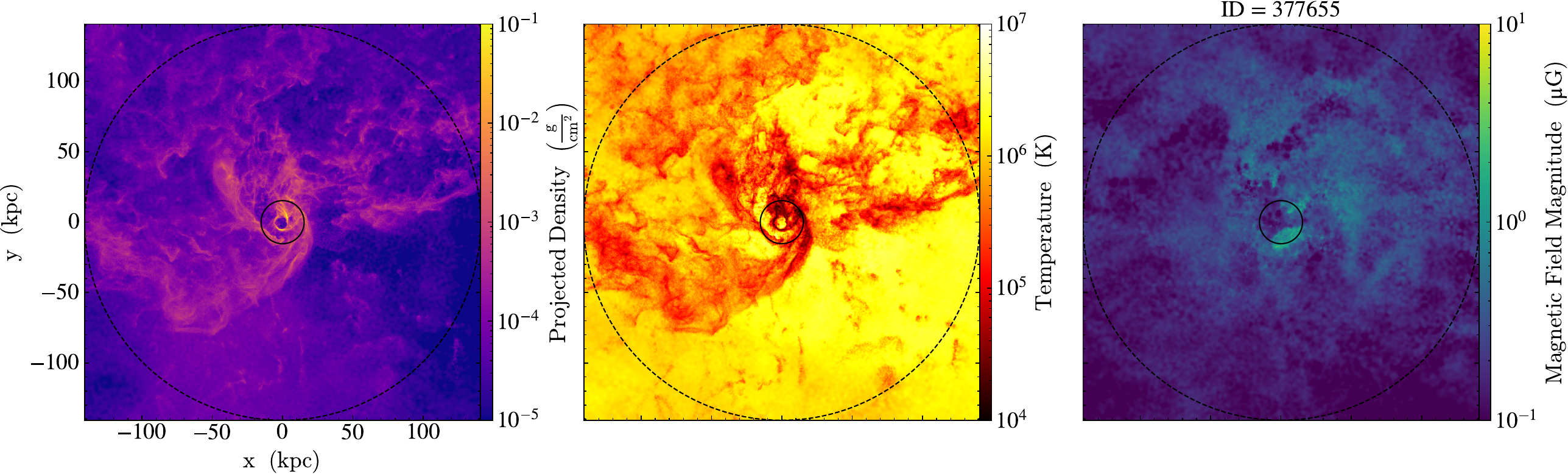}
	\includegraphics[scale=0.4]{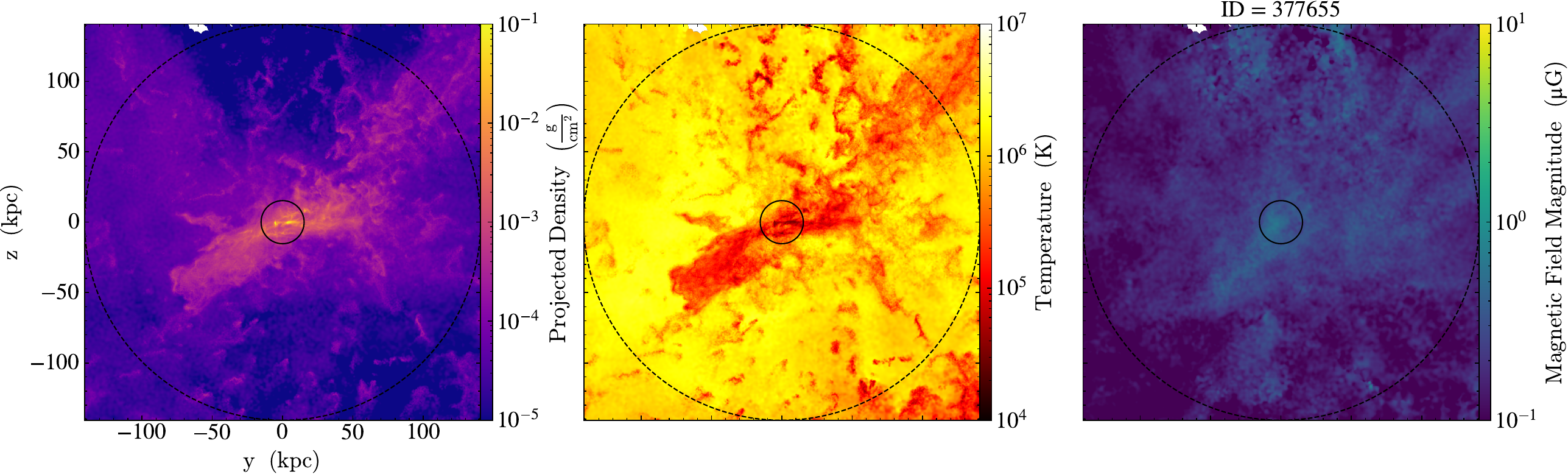}
	\caption{Projected density (left-hand panels), density-weighted temperature (middle panels) and volume-weighted magnetic field (right-hand panels), for two example galaxies in the TNG50 sample. Top and bottom rows for each galaxy represent the face- and edge-on views of a star-forming (ID = 455291, $\Msre \simeq 10^{10.75}$\msun, $\log_{10}\rm{SFR}\simeq-9.93$) and a quenched galaxy (ID = 377655, $\Msre \simeq 10^{10.94}$\msun, $\log_{10}\rm{SFR}\simeq-11.24$), respectively. The inner and outer circles show the $\re$ and 80\% total half-mass, respectively.}
	\label{fig:galaxies}
\end{figure*}
\subsection{Magnetic field treatment in the TNG}\label{sec:B_in_TNG}
One of the most important physical topics added to the TNG in comparison to its successor the Illustris simulation \citep{Vogelsberger2014} is the MHD in its ideal formalism \citep{Pakmor2011,Pakmor2013}.
In the \textsc{arepo}, the divergence constraint is maintained using the eight-wave Powell cleaning scheme \citep{Powell1999}. To deal with the initial condition for the magnetic field in the TNG suite, an initial uniform field in an arbitrary direction with strength of $10^{-8}\,\rm \mu G$ (comoving) is assumed. It has been demonstrated that in both large-scale cosmological \citep{Marinacci2015,Marinacci2016} and zoom simulations \citep{Pakmor2014,Pakmor2017,Garaldi2021}, the initial strength of the magnetic field has a small effect on the final results over several orders of magnitude. This magnetic field is then amplified to its present value as a result of  flux freezing in a condensing gas and different dynamo mechanisms. It should be noticed that in the TNG model, the underlying complex physics of star formation and its non-trivial mutual relation with the magnetic field is not implemented, although the resolution of TNG50 could be comparable to the size of star-forming molecular clouds (MCs).

More specifically, it is argued that at the density and temperature of star-forming regions, the magnetic flux freezing is not a good approximation anymore \citep{Kunz2004} and ambipolar diffusion or turbulent reconnection can dissipate the magnetic field \citep[e.g.,][]{Hosseinirad2018}. However, the turbulent flow arising from cosmic rays, supernova explosions or stellar winds can amplify the magnetic field in what is called small-scale dynamo \citep[e.g.,][]{Hanasz2009}.
\subsection{Galaxy sample}\label{sec:galaxy_sample}
\subsubsection{Finding disc galaxies}
Our first step in this work, is to find disc galaxies in the TNG50 box. Moreover, we select our sample (disc galaxies) from the group catalog and restrict our analysis to the resolved galaxies with stellar mass, $\Mstar > 10^8$ \msun for TNG50. We also exclude galaxies with zero gas component or magnetic field and also those with non-cosmological origin. The latter is done by ignoring galaxies with $\mathtt{SubhaloFlag = 0}$ in the group catalog \citep{Nelson2019}. This leaves us with 6541 galaxies. Several ways are suggested in the literature for finding disc shape galaxies \citep[see][for a comparison]{Zhao2020}. Among them, we utilize two widely used criteria to both dynamically and morphologically separate the sample \citep{Peschken2019,Zhou2020}. In the calculation of these two criteria, we used a slightly modified version of the \textsc{Illustris\_Shapes}\footnote{\url{https://github.com/duncandc/Illustris_Shapes}} code for all star particles inside the effective radius $\re$ considered as twice the stellar half-mass radius $\rh$. The origin of coordinate system is taken to be the position of the most bound particle in each galaxy and the total angular momentum vector of galaxy is taken to be along the $\hat{z}$ axis by rotating the galaxies. The first condition that has to be fulfilled is the disc-to-total mass ratio $D/T$ which we considered to be $\geqslant 0.2$. In order to calculate this ratio, one should discriminate between star particles that dynamically belong to the disc component and others. To do that, we use a so-called circularity parameter $\epsilon$ defined for each star particle as $\epsilon= j_z/\jcirc,$
\citep[][but see also \citealp{Abadi2003} for a slightly different definition and \citealp{Marinacci2014c} for a detailed comparison]{Scannapieco2009} where $j_z$ is the specific angular momentum around the galaxy symmetry axis $\hat{z}$, and $\jcirc=r\,v_\mathrm{c}$, where $r$ is the star radial radius and
\begin{equation}\label{eq:vc}
v_\mathrm{c}=\sqrt{\frac{GM(<r)}{r}}
\end{equation}
computed for the total mass (including the mass of DM, star and wind particles plus gas cells) inside $r$. So, $\jcirc$ is the specific total angular momentum of a particle in a circular orbit in the galaxy. We define $D/T$ as the sum of all star particles masses with $\epsilon>0.7$\citep{Marinacci2014c,Zhao2020}. The second condition we impose is that the galaxy must be morphologically flat. We define a flatness parameter $\mathcal{F}$=$M_1/\sqrt{M_2 M_3}$, where $M_1, M_2$ and $M_3$ are the eigenvalues of the mass tensor such that $M_1$ < $M_2$ < $M_3$, so a smaller $\mathcal{F}$ means a flatter galaxy. We keep galaxies with $\mathcal{F} \leqslant 0.7$ in our sample \citep{Genel2015,Zhao2020,Roshan2021a}. Using these two criteria, our sample will have finally 2376 galaxies.
\subsubsection{Identification of centrals}\label{sec:censat}
The selected sample of disc galaxies in the previous section, could be divided into two groups: centrals and satellites. In the cosmological simulations of galaxy formation, the substructures should be identified with special algorithms. In the TNG, the Friends-of-Friends (FoF) method \citep{Davis1985} is used to find the halos (galaxy clusters or group) and the \textsc{subfind} algorithm \citep{Springel2001} to find subhalos (galaxies) on the fly. Within each halo, the most bound subhalo which is typically the most massive one is considered as the central and the others as satellites. Among our disc galaxies identified in the previous section, 1697 galaxies are identified as centrals and the other 679 ones as satellites. In this paper, we investigate only the central disc galaxies -- henceforth referred to as the the \emph{TNG50 sample} for simplicity -- to reduce environmental effects. These effects will be presented in a forthcoming paper. In our TNG50 sample, the minimum number of all bound gas cells in a galaxy is 605 while the median is 156853.
\subsubsection{Definition of isolated galaxies}\label{sec:isolated}
Galaxies can live in diverse environments, some in rich groups or clusters and some in isolation \citep[e.g.,][]{Ferrarese2012}. It is well known that in addition to the galaxies' mass reservoir, their environment and merger with other galaxies, have substantial effects on their evolutionary path \citep{Naab2017}. In this study to better discriminate between the galaxies' dynamic itself and their environment or external effects, we will focus on the first item and leave the other effects as a subject of another forthcoming paper.
Therefore, in addition to the aforementioned criteria, i.e. being central and disc galaxy, we also categorize the obtained galaxy sample according to their relative isolation in the simulation box. Such a selection has been also used in other studies \citep[see e.g.,][]{Grand2017,Kelley2019,Engler2021}. To this end, using a periodic \textsc{Kdtree}\footnote{\url{https://github.com/patvarilly/periodic_kdtree}} built over all galaxies in the simulation box with the total stellar mass  $\Mstar>10^8$\msun, we first find their nearest neighbours. Then we define $\Riso$ for each galaxy as the distance to its first neighbour in the tree. Using this method, we can define a criteria for the relative isolation of our sample in the TNG50 box. For example, galaxies with $\Riso >2$ Mpc, means those that do not have any neighbour in their 2 Mpc distance, or equivalently those that the distance to their first neighbours are larger than 2 Mpc.

\subsection{Cold gas temperature threshold}\label{sec:cold_gas_def}
Star formation takes place in the dense and cold regions in the ISM, where magnetic field plays a key role \citep[e.g.,][]{Krumholz2019}.
On the other hand, the supernova induced small-scale dynamo occurs in the hot gas. However, in the TNG50 simulation, the stellar feedback is implemented via (magneto)hydrodynamically decoupled wind particles, though it also implicitly contributes to the pressure of the ISM via the pressurised ISM model, hence the supernova-driven amplification of magnetic field is underestimated. For the purpose of this study we consider magnetic field properties both in the total gas cells or those which are colder than a temperature threshold. We assume cells with gas temperatures $\Tcg<5\times10^4\kelvin$ as the cold gas in this study \citep[see e.g.,][]{Ramesh2023}. We also study the effect of a lower threshold temperature of $1.3\times10^4\kelvin$ in \cref{sec:different_T}.

Notice that according to the star formation prescription utilized in the TNG model \citep{Springel2003}, stars form when the density of gas cells exceeds $n_H=0.106$ cm$^{-3}$. Below this density, the state of the gas is controlled by hydrodynamics, whereas for the star-forming gas an average over a two-phase (cold + hot) prescription, results in an effective pressure and temperature ($T\gtrsim 10^4\kelvin$). In our analysis, we take this effective temperature as the star-forming gas temperature.
\subsection{Determination of star-forming and quenched galaxies}\label{sec:sf_or_q}
The role of magnetic field in quenching of star formation in galaxies is still a matter of debate \citep[e.g.,][]{Birnboim2015,Su2017,Kortgen2019,Su2019,Su2021,Whitworth2023}. In this paper, for a more clear view of the problem under study, we will use a commonly accepted criterion \citep[e.g.,][]{Sherman2020} for definition of star-forming and quenched galaxies, in such a way that when we refer to the star-forming (quenched) galaxies, we mean those with the specific star formation rate sSFR greater (lower) than the $10^{-11}~\rm{yr}^{-1}$ value. The sSFR is the ratio of the total instantaneous star formation rate (SFR) to the total stellar mass ($M_{\star}$). To compute SFR, we sum up the instantaneous SFR of the all bound gas cells in each galaxy. Comparing the TNG with observations, \citet{Donnari2020} have recently shown that at $z=0$, this simple definition is consistent with other more sophisticated ones.
\section{Results}\label{sec:res}
In this section, we explore the previously defined TNG50 sample of galaxies for probable correlations between their magnetic field and other physical properties. Moreover, we compare the simulation results with what \taba found for their sample of non-interacting non-cluster nearby galaxies. In the following, we take $B$ as the total magnetic field strength (or simply the magnetic field), unless otherwise explicitly noted\footnote{In \cref{sec:compare_with_T16}, we use $\btot$ to differentiate between the square root of the volume weighted value of $B^2$ and its ordered and turbulent components.}. We calculate the square root of the volume-weighted $B^2$ of a galaxy as 
\begin{equation}\label{eq:bbar}
	\bbar=\sqrt{\dfrac{\sum_{i=1}^{n} B_i^2 V_i}{\sum_{i=1}^n V_i}}.
\end{equation}
The summation is over the magnetic field and volume of gas cells within a specific volume in the galaxy.

Using a kernel density estimate from \textsc{seaborn} python package\footnote{\url{https://seaborn.pydata.org/generated/seaborn.kdeplot.html}}, in \cref{fig:kde_density_plot_B} we show the probability density functions (pdf) of $\bre$ for different samples. For each galaxy, $\bre$ denotes to $\bbar$ calculated in a sphere with the centre at its most bound particle and the radius of $\re$, the galaxy effective radius. The figure compares the magnetic field distribution of all the central galaxies (including disc and non-disc galaxies with non-zero magnetic fields) with the total stellar mass $\Mstar>10^8$\msun in the simulation box against the TNG50 sample (which includes only the central disc galaxies with the same total stellar mass limit as the former sample). The small vertical lines mark the median values of the corresponding distributions. The red dashed curve, shows $\bre$ for the first sample. This distribution shows a tail towards the larger magnetic field values with a median $\simeq 4.13\,\mu$G . For the TNG50 sample (the orange dotted curve; 1697 galaxies), this tail is removed and the distribution shows a Gaussian shape with a decreased median $\simeq 3.50\,\mu$G. Constraining the TNG50 sample to the one that includes only the isolated galaxies with $\Riso>2$Mpc (the green dash-dotted curve; 103 galaxies) does not change the distribution, just a tiny ($\simeq 0.18\,\mu$G) decrease in the median value and the density around it. When we consider magnetic fields only in the cold gas cells in the last sample, the dispersion of the distribution is decreased whereas its median remains almost unchanged.

\cref{fig:galaxies} illustrates the face- and edge-on views of two typical galaxies in the TNG50 sample that are identified as a star-forming (ID = 455291, $\Msre \simeq 10^{10.75}$\msun, $\log_{10}\rm{SFR}\simeq-9.93$) and a quenched galaxy (ID = 377655, $\Msre \simeq 10^{10.94}$\msun, $\log_{10}\rm{SFR}\simeq-11.24$) using the methods introduced in \cref{sec:sf_or_q}. $\Msre$ means the total stellar mass inside the effective radius $\re$. For each galaxy, the left, middle and right columns show the projected density, the density weighted temperature and the volume weighted magnetic field. The inner circle has a radius equal to $\re$ while the outer circle radius is 0.8 of the total half-mass radius. Although the objective of this figure is rather illustrative, we can observe a tight relationship between the high density gas structure and its lower temperature. Moreover, the overall structure of the magnetic field follows the denser and cooler patterns in both galaxies, but due to the volume weighted visualization of the magnetic field, more weight is assigned to the larger low density cells. When viewed face-on, a remarkable ring-like feature is observed at the centre of both galaxies where the density and the magnetic field have lower values comparing with their immediate surroundings. This could be related to the activity of AGN that pushes gas out of the centre of galaxies. Such a feature is also reported by \citet{Whittingham2021} (their 1605-3M model) in their zoom-in MHD simulations where they used the same implementation of MHD in \textsc{AREPO} code and an AGN feedback in the framework of Auriga galaxy formation model \citep{Grand2017}.
%===============
\subsection{Magnetic field Correlations in the TNG}\label{sec:B_corr_TNG}
\begin{table}
	\centering
	\caption{Correlation of $\bre$, $\bcgre$, $\bd$ and $\bcgd$ with $\vre$ and $\Mtotre$ for galaxies in the TNG50 sample. The subscripts $\rm{\re}$ and $\rm{d}$ denote whether the average is calculated over a sphere or a disc with the radius $\re$. The disc has a width of 1 kpc. The subscript $\rm{cg}$ means that only the colder gas cells with $T < 5\times10^4$ are taken into account. In the 3rd column, ``n" and ``y" denote to the samples with no isolation and $\Riso>2$ Mpc, respectively. The next two columns are the Pearson correlation coefficient ($\Pearson$) and the slope of a least squares fitting ($a$). Same quantities are also reported in the two last columns ($\Pearsonsf$ and $a_{\rm{sf}}$) by restricting the samples with no isolation to those which include only the star-forming galaxies i.e. $\Pearsonsf$ and $a_{\rm{sf}}$. Since the isolated samples do not include any quenched galaxy, $\Pearsonsf$ and $a_{\rm{sf}}$ are equal to the previous columns for these samples and we do not repeat them.}
	\label{tab:vre_and_Mtot_vs_B}
	\begin{tabular}{ccccccc}
		\hline
		Property & $\bbar$ & isolation & $\Pearson$ & $a$ & $\Pearsonsf$ & $a_{\rm{sf}}$ \\
		\hline
		$\vre$ & $\bre$      & no & 0.09 & 0.14 & 0.27 & 0.39  \\%
		...    & $\bcgre$    & ... & 0.35 & 0.47 & 0.46 & 0.63  \\%
		...    & $\bd$       & ... & 0.23 & 0.34 & 0.46 & 0.62  \\%
		...    & $\bcgd$     & ... & 0.38 & 0.50 & 0.50 & 0.66 \\%
		\\
		$\vre$ & $\bre$      & yes & 0.49 & 0.77 & -    & -     \\%
		...    & $\bcgre$    & ... & 0.59 & 0.91 & -    & -     \\%
		...    & $\bd$       & ... & 0.64 & 1.01 & -    & -     \\%
		...    & $\bcgd$     & ... & 0.65 & 1.00 & -    & -     \\%
		\\
		$\Mtotre$  & $\bre$  & no & -0.11 & -0.05 & -0.03 & -0.01  \\%
		...    & $\bcgre$    & ... & 0.12  & 0.05  & 0.16  & 0.07  \\%
		...    & $\bd$       & ... & 0.08  & 0.04  & 0.20  & 0.08  \\%
		...    & $\bcgd$     & ... & 0.17  & 0.07  & 0.23  & 0.10  \\%
		\\
		$\Mtotre$  & $\bre$  & yes & 0.21  & 0.10  & -     & -     \\%
		...    & $\bcgre$    & ... & 0.33  & 0.15  & -     & -     \\%
		...    & $\bd$       & ... & 0.41  & 0.19  & -     & -     \\%
		...    & $\bcgd$     & ... & 0.41  & 0.19  & -     & -     \\%
		\hline
	\end{tabular}
\end{table}

\begin{figure*}
	\centering
	\includegraphics[scale=0.52]{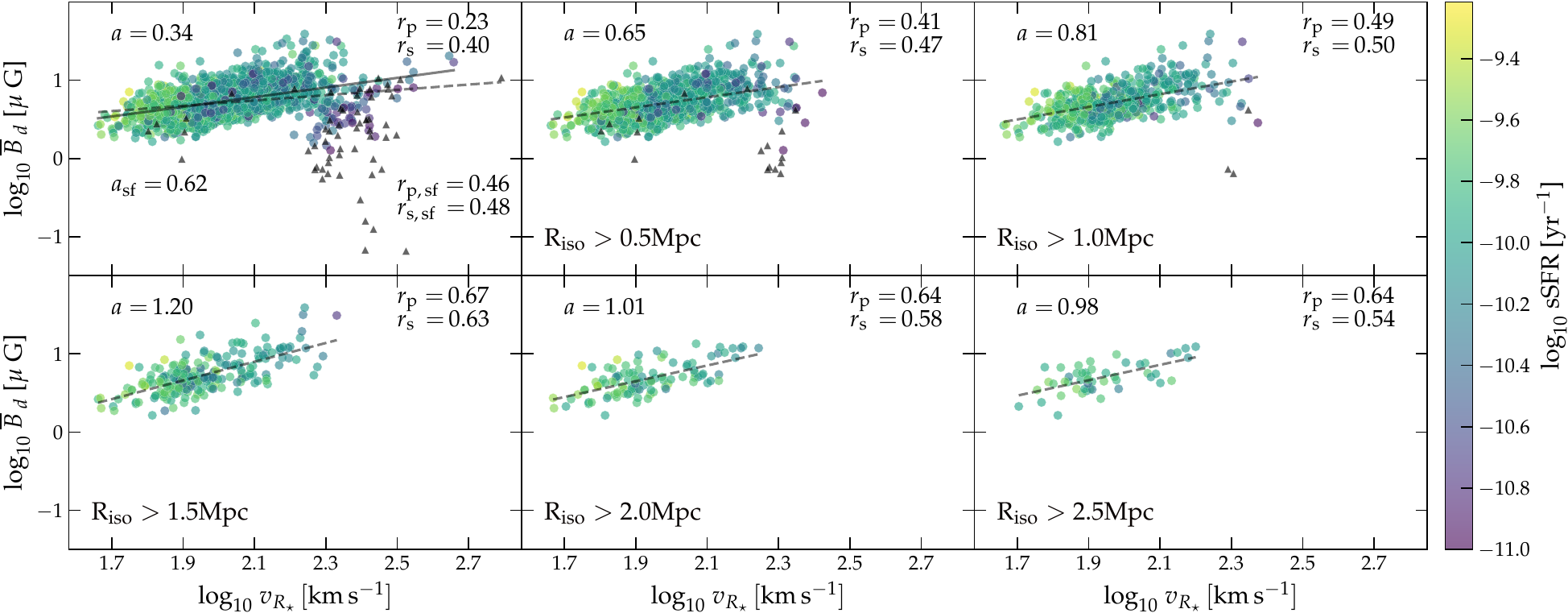}
	\caption{$\bd$ vs. $\vre$ for the TNG50 sample galaxies. The color coding shows sSFR. Panels illustrate different isolation criteria as explained in the text. The circles denote to the star-forming galaxies whereas the black triangles denote the quenched ones with sSFR $<10^{-11}$. In each panel, $a$ is the slope of a least squares fit over all the demonstrated galaxies (shown by a dashed-line). Likewise, $\Pearson$ is the Pearson correlation coefficient and $\Spearman$ is the Spearman rank coefficient that are calculated for all the galaxies in that panel. In the upper-left panel, $\asf$ (the solid line slope), $\Pearsonsf$ and $\Spearmansf$ are the calculated corresponding quantities for the star-forming galaxies.
    }
	\label{fig:Bd_Vre_sSFR}
\end{figure*}
\begin{figure*}
	\centering
	\includegraphics[scale=0.52]{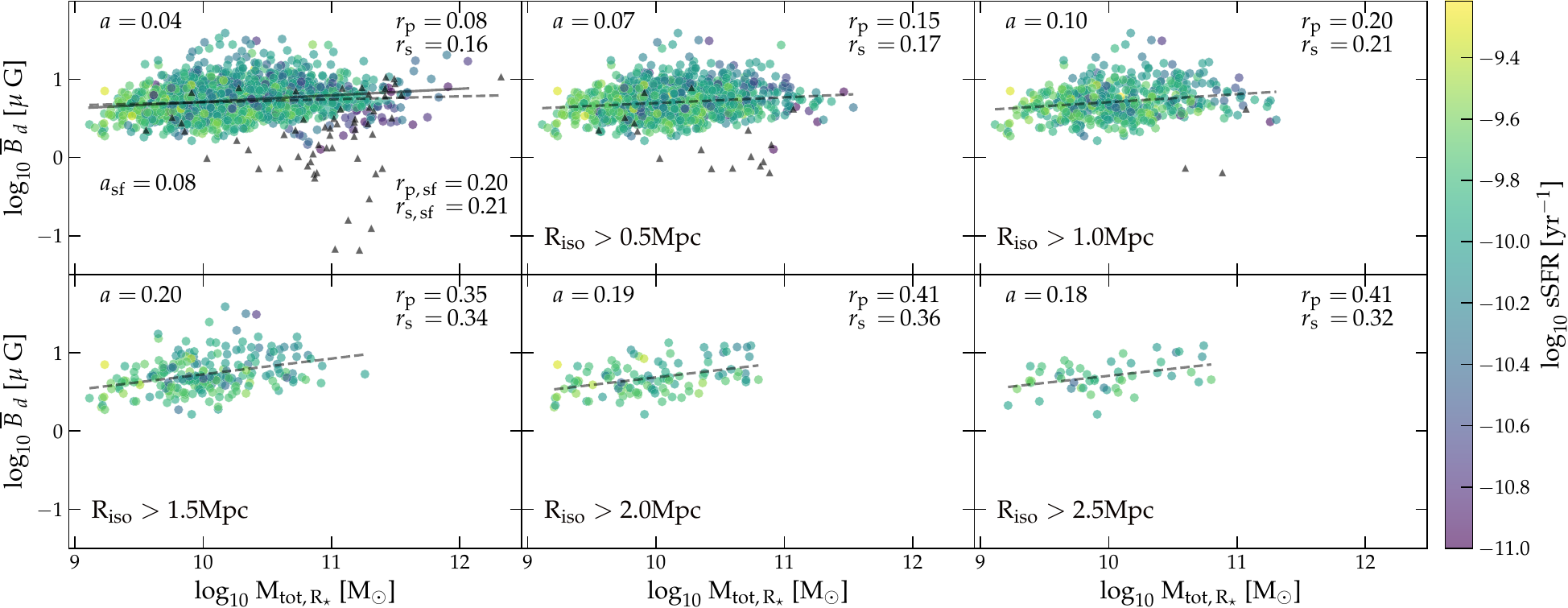}
	\caption{Same as the \cref{fig:Bd_Mre_sSFR}, but here $\bd$ vs. $\Mtotre$ is shown. $\Mtotre$ is the total mass of a galaxy inside its effective radius $\re$.}
	\label{fig:Bd_Mre_sSFR}
\end{figure*}
%============= minipage example
\begin{figure}
	\includegraphics[width=\linewidth]{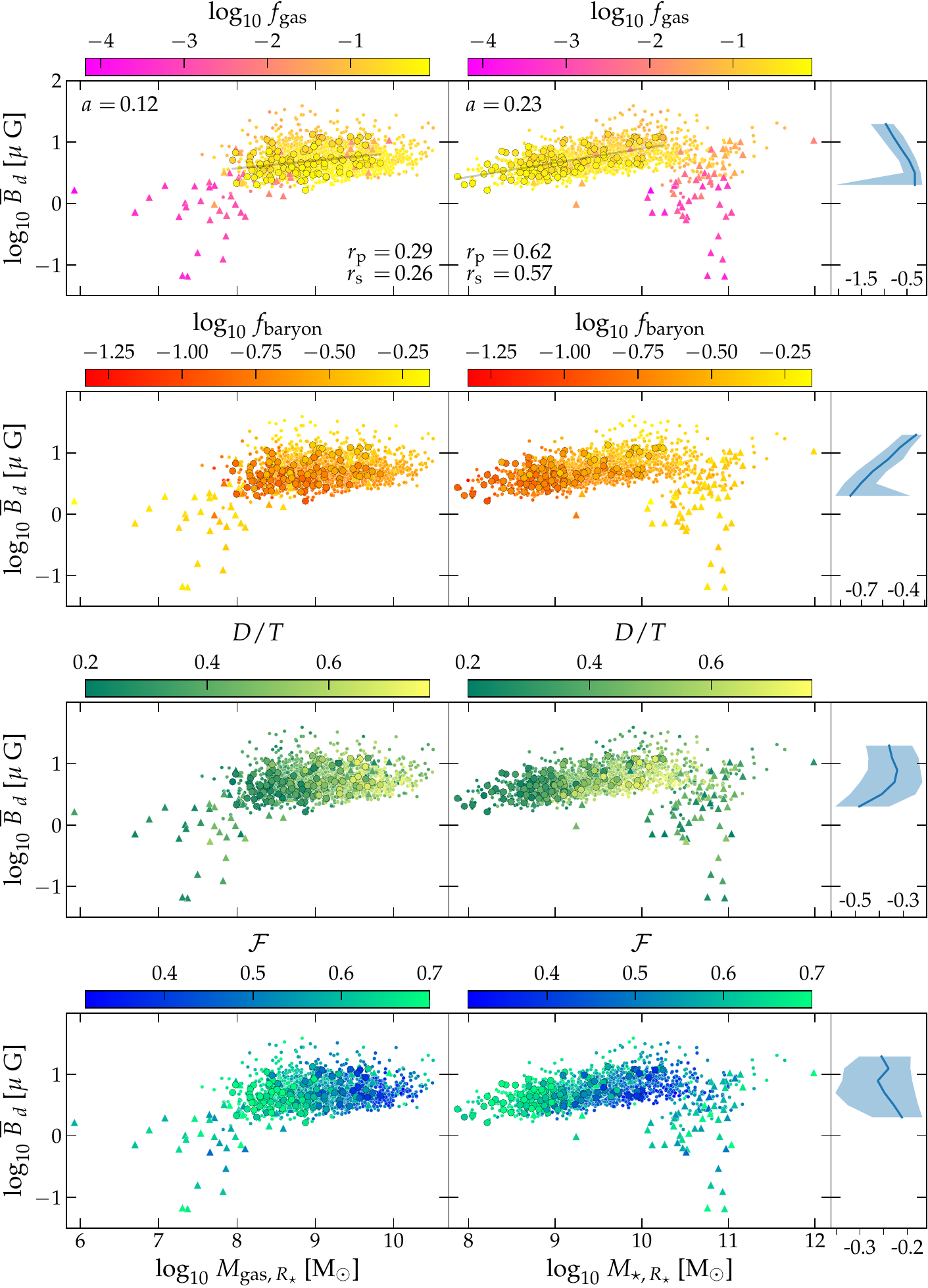}
	\caption{$\bd$ vs. the total gas mass (\textbf{left panels}) and the stellar mass (\textbf{middle panels}), both within the effective radius ($\re$), for the star-forming and the quenched galaxies in the TNG50 sample (small circles and triangles, respectively) and those with $\Riso>2$ Mpc (large circles). The colour bars visualise from top to bottom $\fgas$, $\fbar$, $D/T$ and $\mathcal{F}$, respectively. The slope of the least squares fit and the correlation coefficients for the isolated sample are written in the upper panels where the solid lines show the the least square fit. \textbf{Right panels} show from top to bottom the median of $\log_{10}\bd$ (blue solid curve) vs. $\log_{10}\fgas$, $\log_{10}\fbar$, $\log_{10}D/T$ and $\log_{10}\mathcal{F}$, respectively. The shaded area shows the 16th-84th percentiles. We do not show medians for bins with data points $< 15$.}
	\label{fig:Bd_Mgasre_Mstarre_cen}
\end{figure}
\begin{figure*}
	\centering
	\includegraphics[scale=0.40]{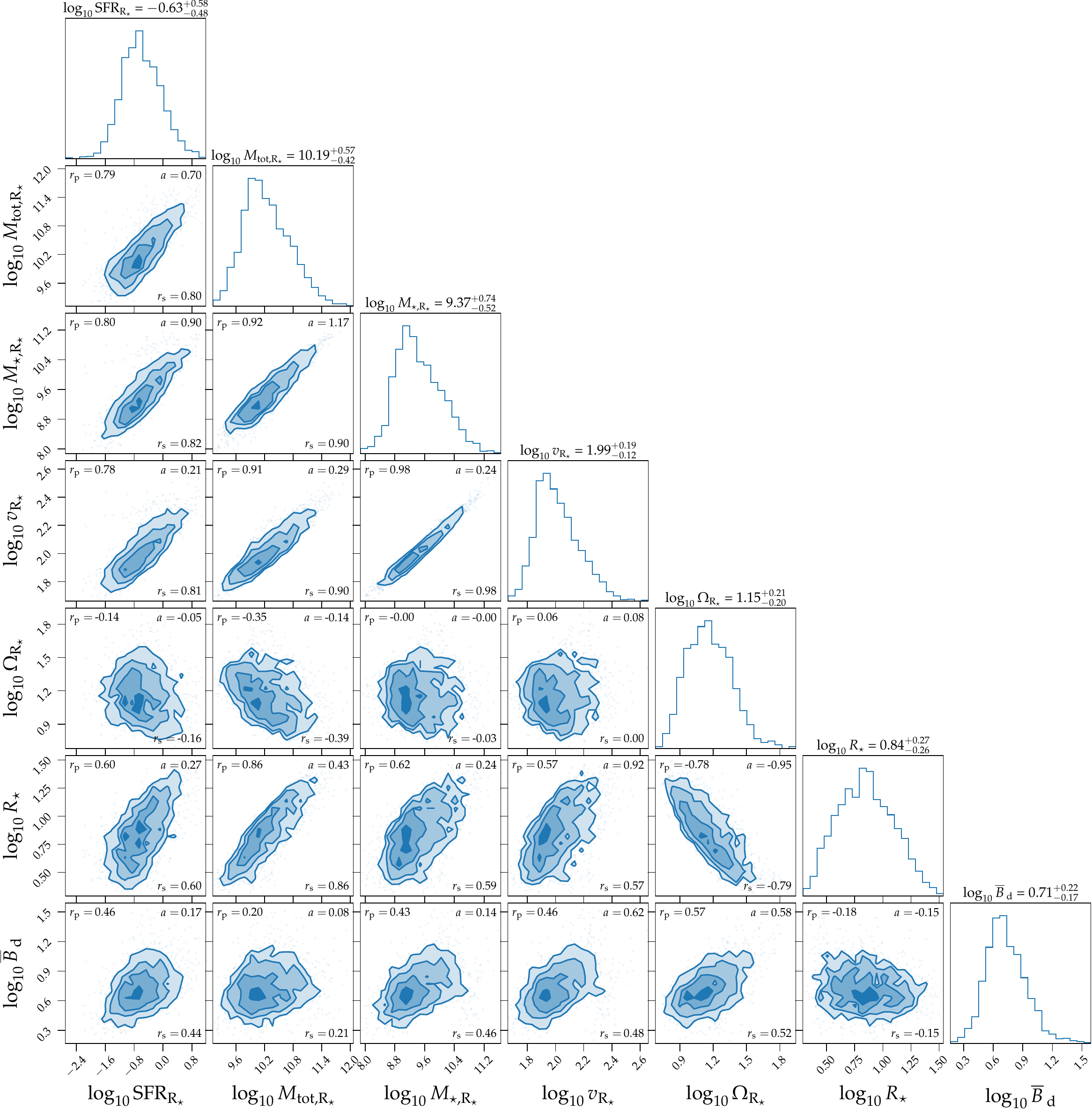}
	\caption{Correlation of different properties of the star-forming galaxies in the TNG50 sample, without isolation criterion, with respect to each other. Units are: \msunyr for $\sfrre$, \msun for $\Mtotre$ and $\Msre$, \kms for $\vre$, \kms\,kpc$^{-1}$ for $\Omegare$, kpc for $\re$ and $\mu$G for $\bd$.}
	\label{fig:corner_cent_Bd}
\end{figure*}
\begin{figure*}
	\centering
	\includegraphics[scale=0.45]{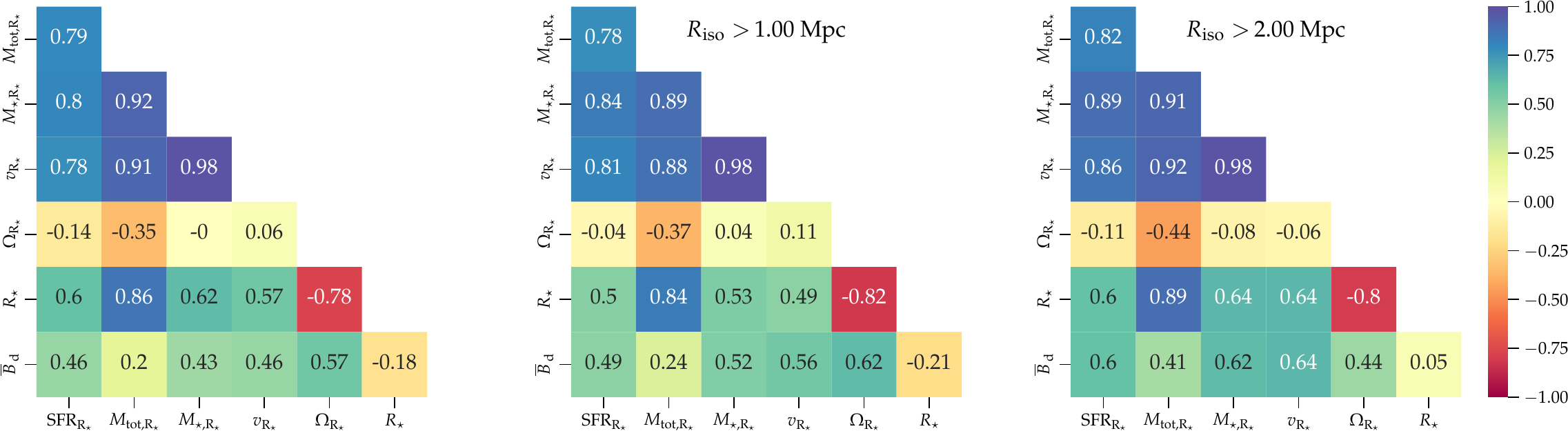}
	\caption{The Pearson correlation coefficient ($\Pearson$) between various properties of the star-forming galaxies in the TNG50 sample, with no isolation criterion (left panel), $\Riso>1$Mpc (middle panel) and $\Riso>2$Mpc (right panel).
	}
	\label{fig:grid_corrs_cen_iso_Bd}
\end{figure*}
\subsubsection{Correlation with rotational velocity and mass}\label{sec:B_v_mass_corr_TNG}
Having identified the TNG50 sample, we can study the correlation between the magnetic field, total mass and rotational velocity. To this end, we calculate $\bbar$ for the gas cells within a disc $\bd$ for each galaxy according to \cref{eq:bbar}. This disc is defined as a cylinder with the radius of $\re$ and the height of 1 kpc, centred on the galaxy most bound particle.As mentioned in \cref{sec:cold_gas_def}, we can also consider a temperature cut for selecting gas cells, to study the relations in colder gas regions. The rotational velocity at $\re$ is calculated as $\vre = \sqrt{G\Mtotre/\re}$ where $\Mtotre$ is the total mass (DM + stars + gas) of a galaxy within $\re$.

We list the Pearson correlations $(\Pearson)$ of $\bd$ with $\vre$ and $\Mtotre$ along with the least squares fit slopes $(a)$ for $\bd-\re$ and $\bd-\Mtotre$ relations in \cref{tab:vre_and_Mtot_vs_B}. For comparison, the above analysis is also repeated for $\bre$. The results for star-forming galaxies and isolated galaxies ($\Riso>2$ Mpc) in the TNG50 sample are also separately listed. \cref{tab:vre_and_Mtot_vs_B} shows that whereas $\bre$ and $\bd$ are not correlated with $\vre$ and $\Mtotre$ in the TNG50 sample, the average magnetic fields in the colder gas cells ($\bcgre$ and $\bcgd$) show weak correlation with $\vre$ ($\Pearson=0.35$ and 0.38, respectively), however $\Mtotre$ is still not correlated. The star-forming galaxies correlation coefficients $(\Pearsonsf)$ of $\bbar$ with $\vre$ and $\Mtotre$ in the TNG50 sample are stronger than those of the complete sample which includes the quenched galaxies as well. Their slope $\asf$, are also steeper. This increase is more pronounced for $\bre$ and $\bd$, with respect to $\bcgre$ and $\bcgd$. This can be due to the fact that the faster-rotating/more massive central galaxies have weaker magnetic fields in their mostly consisted of warmer gas (we will discuss it in more details in \cref{sec:BH}). Interestingly, it can also be seen from \cref{tab:vre_and_Mtot_vs_B} that restricting the TNG50 sample galaxies to the more isolated ones has, generally, a positive effect on $\bbar-\vre$ and $\bbar-\Mtotre$ correlations. Furthermore, $\bbar$ in the colder gas displays tighter correlations as a whole. Finally, we note that comparing the equivalent correlations for the the disc and the sphere, the former are stronger. As a consequence, the tightest relation of $\bbar$ with $\vre$ and $\Mtotre$ is observed for $\bcgd$ in the isolated sample. Notice that these galaxies are all star-forming too.

Here, with the aim of better understanding of the existent correlations, we study $\bd$ relations in more details. In this regard \cref{fig:Bd_Vre_sSFR} demonstrates the coupling between $\bd$ and $\vre$. In this figure, the upper-left panel shows the TNG50 complete sample while in the others we impose the isolation criterion. From left to right and top to down, the minimum value of $\Riso$ is increased from 0.5 Mpc to 2.5 Mpc. The color-bar shows sSFR. In each panel, we also show the fit slope $a$, $\Pearson$ and also the Spearman rank coefficient $(\Spearman)$. In the upper-left panel, $\asf$, $\Pearsonsf$ and $\Spearmansf$ are the corresponding values for the star-forming galaxies. We can see that $\bd$ distribution spans a range from $\sim 0.06\,\mu$G (in quenched galaxies) to $\sim 38.96\,\mu$G with the median $\sim 5.07\,\mu$G. Regardless of the isolation criterion, $\bd$ is correlated with $\vre$ $(a\geqslant 0.34$, $\Pearson \geqslant 0.23$ and $\Spearman \geqslant 0.40)$. The figure also exhibits that most of the quenched galaxies and those with the lowest sSFR, are placed under the fitted line and hence have relatively smaller $\bd$ compared to the galaxies with the same $\vre$. By increasing the isolation radius, galaxies with lower sSFR values are removed gradually from the TNG50 sample which means these galaxies live preferentially in more involved regions. More specifically, the correlation becomes super-linear $(a=1.20)$ for the sample with $\Riso>1.5$ Mpc.

\cref{fig:Bd_Mre_sSFR} shows the correlation of $\bd$ with $\Mtotre$. Same as the \cref{fig:Bd_Vre_sSFR}, the correlations gently become stronger with increasing isolation, but unlike $\vre$, the maximum of $\Pearson$ and $\Spearman$ are observed for galaxies with $\Riso>2$ Mpc, where they reach to 0.41 and 0.36, respectively. Moreover, the slope of the linear fit also increases from 0.04 for the complete sample to 0.20 for the galaxies that meet the $\Riso>1.5$ Mpc criterion, though it then sees a decline. Comparing \cref{fig:Bd_Vre_sSFR,fig:Bd_Mre_sSFR}, we find that the $\bcgd-\vre$ correlation is tighter than the $\bcgd-\Mtotre$ one. Moreover, the treatment of the quenched galaxies seems almost the same. In addition, as already mentioned, excluding the quenched galaxies from the TNG50 sample reveals a modest increase in the correlation and the slope values (see the upper-left panels in both of the figures).

In \cref{fig:Bd_Mgasre_Mstarre_cen}, we demonstrate $\bd$ vs. the total gas (cold + non-cold, left panel) and the stellar mass components (middle panel) both within $\re$ for the TNG50 sample galaxies (small circles) in which those that satisfies the $\Riso>2$ Mpc criterion are also denoted (larger circles). In the upper left and middle panels, the labels shows $\Pearson$, $\Spearman$ and $a$, for the $\bd-\Mgre$ and $\bd-\Mstar$ relations in the isolated sample, respectively. We can see that the stellar component correlations with $\bd$ are $\Pearson=0.62$ and $\Spearman=0.57$, with a best-fit slope of $a=0.23$. The gaseous matter has weaker correlation ($\Pearson=0.29$ and $\Spearman=0.27$). Comparing these values with ones for $\Mtotre$ in the lower middle panel of \cref{fig:Bd_Mre_sSFR}, we find that $\Msre$ has the tightest correlation. Because of the magnetic flux-freezing, the strength of the magnetic field pervading the gas is stronger in these denser and colder regions. Hence, the average magnetic field is expected to correlate positively with the stellar mass in star-forming galaxies of the TNG model. It should be noted that although with the TNG50 high resolution, the gas flow is reasonably simulated well, the embedded star formation prescription does not include explicit supernova feedback. Thus the turbulent nature of the ISM and its effect on the magnetic field is probably underestimated, though it seems rather unclear how much supernovae actually contribute to the overall turbulence in disc galaxies \citep[see e.g.,][for recent studies]{Pfrommer2022,Bieri2022}.

To better comprehend other probable underlying correlations, we have also colour-coded the circles to reflect four other quantities related to each galaxy. This includes from top to bottom in each panel the gas fraction $\fgas=\Mgre/(\Mgre+\Msre)$, the baryon fraction $\fbar=(\Mgre+\Msre)/\Mtotre$, the disc-to-total mass ratio $D/T$ and the flatness parameter $\mathcal{F}$, respectively. Looking at the left panels, one can see that among the star-forming galaxies with equal gas content, generally those with lower gas fraction have relatively larger $\bd$ values. This also holds for $\fbar$ as well but in reverse. Furthermore, for the star-forming ones with equal stellar mass, no correlation holds between $\bd$ with $\fgas$ and $\fbar$. The quenched galaxies span a wide range of $\Mgre$ ($10^6-10^{10}$ \msun) and relatively a smaller range of $\Msre$. For these galaxies, those with larger $\fgas$ have relatively stronger $\bd$, however, for the $\fbar$ we can not see such a trend. We should add that the other two properties, i.e. $D/T$ and $\mathcal{F}$ do not show any connection with $\bd$ for galaxies with equal gaseous or stellar masses.

The right panels exhibit the general trend of $\bd$ (vertical axis) with respect to the same four features (horizontal axes). We plot the median (blue solid) and 16th and 84th percentiles (shaded region). The medians in the two upper panels display inverse trends. With decreasing $\bd$, $\fgas$ increases with a peak around $2\,\mu$G, whereas, $\fbar$ decreases with the lowest value at the same magnetic field strength. With regards to the other two quantities, they do not monotonically decreasing or increasing.
%============
\subsubsection{Correlations of magnetic field with other galaxy properties}\label{sec:corr_with_others}
In the previous section we studied the correlation of the magnetic field with the rotational velocity and different mass components of galaxies in the TNG50 sample. It is worthwhile to see how these galaxy features plus a few other ones introduced in the following, correlate with the magnetic field as well as with each other. This could help us to find the more basic correlations. Here, we restrict our analysis to the star-forming galaxies in the TNG50 sample, as we already observed that the quenched galaxies deviates from the main relation.

In \cref{fig:corner_cent_Bd} the corner plot displays the correlation of some of the properties of the TNG50 galaxy sample, namely $\bd$, $\sfrre$, $\Mtotre$, $\Msre$, $\vre$, $\re$ and  $\Omegare$ with respect to each other. The final property is the angular velocity at the effective radius $\Omegare=\vre/\re$. One should keep in mind that the underlying star formation process in the TNG galaxy formation model is explicitly tuned to reproduce the empirical Kennicutt-Schmidt relation \citep[$\Sigma_{\rm{SFR}}\sim\Sigma_{\rm{gas}}^{1.4}$,][]{Kennicutt1998,Springel2003}. So, this relation is not a prediction of the simulation.

We can see from \cref{fig:corner_cent_Bd} that the correlations of $\sfrre$ with $\Msre$, $\Mtotre$ and $\vre$ are all strong ($\Pearson\geqslant0.78$). It is obvious that the tightest correlation is $\Msre-\vre$ with $\Pearson=0.98$. It is also remarkable that $\re$ correlates tightly with $\Mtotre$ ($\Pearson=0.86$) but not tightly with other properties, though it shows also a tight inverse correlation with $\Omegare$ ($\Pearson=-0.78$). We can, however, observe that the correlation between $\bd$ and $\Mtotre$, $\Msre$, $\vre$, $\re$ and $\sfrre$ are all weaker than the correlation of these properties with each other. More specifically, the correlation of $\vre$ (or equivalently $\Msre$, as they are tightly correlated) with $\sfrre$ is stronger than the $\bd-\vre$ and $\bd-\sfrre$ correlations. As the two latter correlations are nearly the same, one can hardly say which one is more original, although the first is a little bit stronger (the difference is more pronounced for the more isolated sample; see \cref{fig:grid_corrs_cen_iso_Bd}). In \taba, the authors found that the $\SI-\vrot$ correlation with $\Spearman=0.72$ could be induced by a more original correlation with SFR ($\Spearman\sim0.9$), as SFR and $\vrot$ are in fact correlated ($\Spearman=0.67$) with each other. However, for the $\SPI-\vrot$ correlation with $\Spearman=0.80$, they concluded that this correlation should be more direct than the less tight correlation of SFR with $\vrot$.

In \cref{fig:grid_corrs_cen_iso_Bd}, the effect of isolation on the aforementioned correlations is demonstrated. The left panel shows $\Pearson$ for the same sample, i.e. no isolation criterion is imposed, whereas in the middle and right panel our sample includes galaxies with $\Riso>1$ Mpc and $\Riso>2$ Mpc, respectively. This figure shows that galaxies in more isolated environments exhibit slightly tighter correlations between their $\sfrre$, $\Mtotre$, $\Msre$, $\vre$ and $\re$ as a whole. For $\bd$, this increase in $\Pearson$ is greater. This indicates that the correlation of the magnetic field with $\sfrre$, $\Mtotre$, $\Msre$ and $\vre$ is sensitive to the isolation or the environment.
%=========================================
\section{Discussion}\label{sec:discussion}
%============
\begin{figure*}
	\centering
	\includegraphics[width=0.85\linewidth]{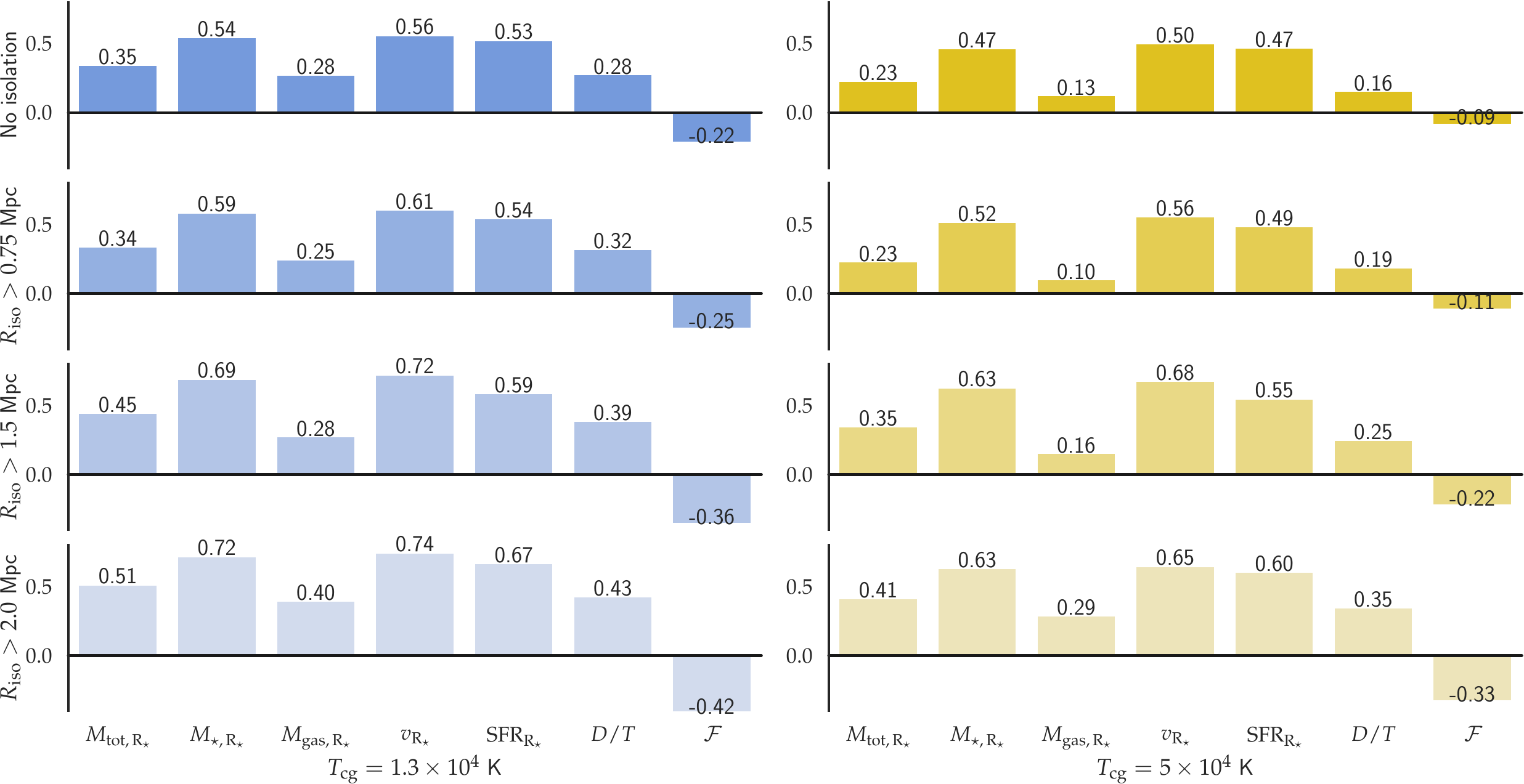}
	\caption{The Pearson Correlation coefficient between $\bcgd$ and other properties of star-forming galaxies in the TNG50 sample (horizontal axis). From top to bottom, the $\Riso = 0$ (all galaxies), $\Riso>0.75$ Mpc and $\Riso>1.5$ Mpc, respectively. Left: $\Tcg=10^4\kelvin$ and right: $\Tcg=5\times10^4\kelvin$.}
	\label{fig:bar_corrs_cen_iso_T13000}
\end{figure*}
\subsection{Effect of different temperature thresholds}\label{sec:different_T}
In previous sections, we calculated the volume weighted magnetic field by taking the average over ``all" the gas cells in a disc or sphere. In \cref{tab:vre_and_Mtot_vs_B}, however, we saw that the correlation between the magnetic field in the cold gas phase (either $\bcgd$ or $\bcgre$) with $\Tcg<5\times10^4$ and $\vre$ or $\Mtotre$ is stronger than the correlation of the magnetic field in all the gas. We expect that decreasing this temperature threshold should possibly make the correlation even stronger.

Here we employ an extra lower temperature threshold $\Tcg=1.3\times10^4\kelvin$ to find the correlation between $\bcgd$ and $\Mtotre$, $\Msre$, $\Mgre$, $\vre$ with the additional properties $\sfrre$, $D/T$ and $\mathcal{F}$ also included. We plot \cref{fig:bar_corrs_cen_iso_T13000} to test this idea in terms of the temperature threshold and also the isolation radius. In the figure, the left and right panels have $\Tcg=1.3\times10^4\kelvin$ and $\Tcg=5\times10^4\kelvin$, respectively and from top to bottom the minimum isolation radius is increased. Comparing the correlations with regards to the gas temperature, we can see that apart from $\mathcal{F}$, other galaxies' properties exhibit weaker correlations for the higher gas temperature threshold. In the case of $\mathcal{F}$, it shows an anti-correlation that becomes weaker considering the higher $\Tcg$. It is also interesting that for both temperatures, the correlation (for $\mathcal{F}$ the anti-correlation) of $\bcgd$ with all other quantities becomes tighter with increasing the isolation. 
%============
\begin{figure}
	\centering
	\includegraphics[scale=0.3]{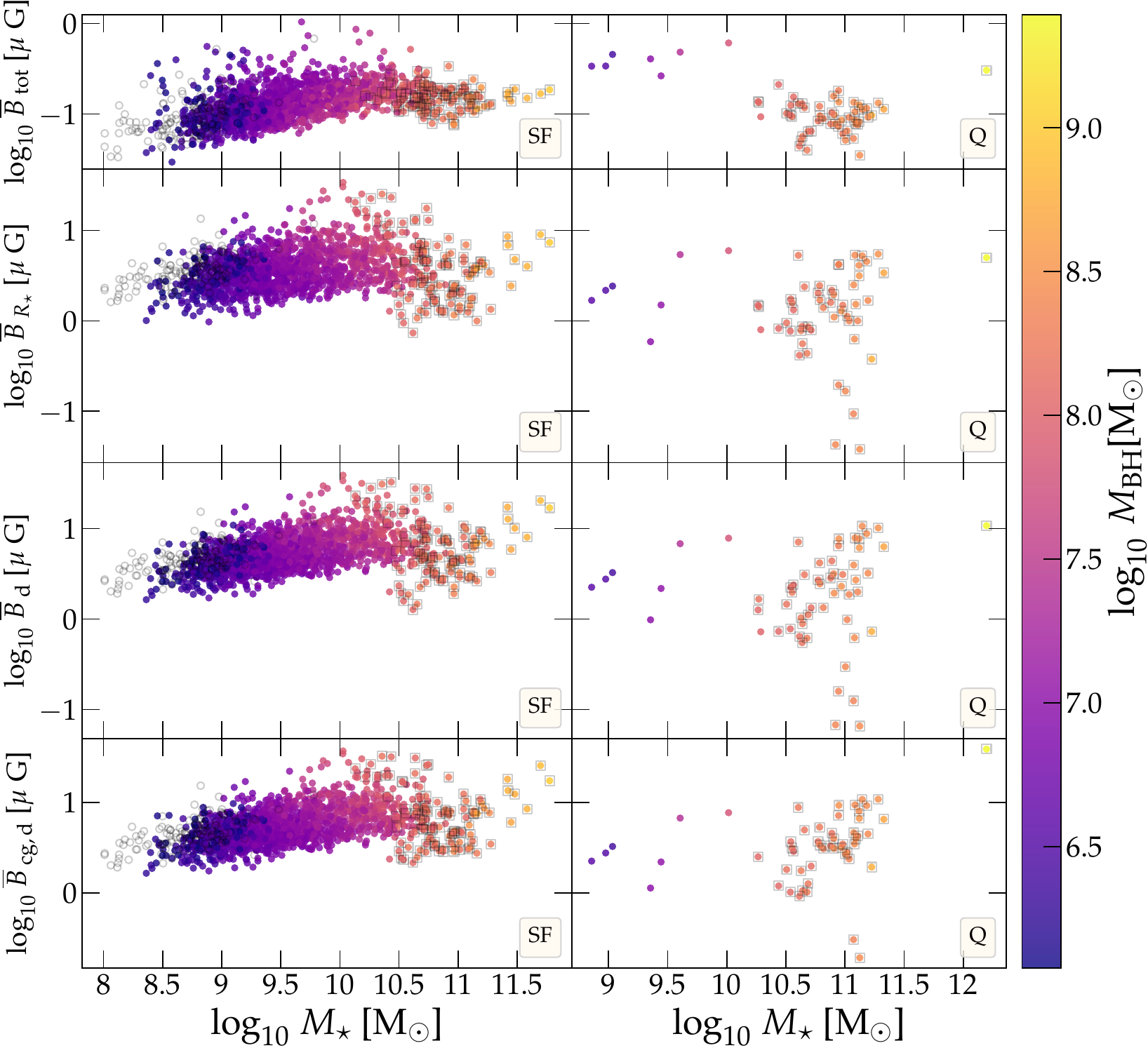}
	\caption{$\overline{B}$ vs. total stellar mass. Galaxies in the TNG50 sample are depicted separately in two groups. \textbf{Left:} Star-forming galaxies with sSFR $\geqslant 10^{-11}$ yr$^{-1}$. \textbf{Right:} Quenched galaxies with sSFR < $10^{-11}$ yr$^{-1}$. Note that a few quenched galaxies are not shown in the most-bottom panel, as they do not have any cold gas cell in their disc. The color coding shows the black hole mass hosted by each galaxy. The empty circles show galaxies without black hole. Galaxies that host black holes with masses $\Mbh > 10^8$\msun are marked by an extra grey square for more clarity. From top to bottom the vertical axes are: $\btot$, $\bre$, $\bd$ and $\bcgd$.}
	\label{fig:Btot_Mstar_cen_SFQ_MBH}
\end{figure}
\subsection{Effect of AGN feedback on the correlation}\label{sec:BH}
Recent studies suggest that AGN feedback could help galaxy formation models to be in better agreement with observations \citep[e.g][]{Raouf2018,Raouf2019a,Raouf2023,Kondapally2023}. As an example, in massive galaxies star formation could not be quenched by only stellar feedback mechanisms such as stellar winds and supernova explosions \citep{Benson2003,Vogelsberger2020}. In the TNG galaxy formation model, the AGN feedback allows to suppress the star formation in massive galaxies besides helping to rectify other model shortcomings \citep{Xu2022}. This is also argued to be the case in other cosmological galaxy simulations \citep[e.g.,][]{Croton2006,Bower2017,Raouf2017,Katsianis2020,Wells2020b}.
In the TNG model, a two-way prescription for the AGN feedback is utilized. For low-accretion rates, momentum is injected into the surrounding medium via black hole-driven winds in random direction (kinetic mode), while in high-accretion rates only the thermal energy transmission occurred and the surrounding gas temperature is accordingly increased (quasar mode) \citep{Weinberger2017}. Black holes with the masses of $\sim 1.2\times10^6$\msun are seeded and maintained at the centre of potential well of each halo as soon as their total masses exceeds $7\times10^{10}$\msun \citep{Pillepich2018a}. In the TNG framework, it is argued that the kinetic mode is the dominant mechanism for the quenching of massive central galaxies \citep{Donnari2021a}.

Here, to better understand the effect of AGN feedback on the average magnetic field of galaxies in the TNG50 sample, we have plotted \cref{fig:Btot_Mstar_cen_SFQ_MBH} in which from top to bottom $\btot, \bre, \bd$ and $\bcgd$ are plotted vs. the total stellar mass $(\Mstar)$ of these galaxies separately for the star-forming (left panels) and quenched galaxies (right panels). The first three quantities represent the the square root of the volume weighted average of $B^2$ in the total gas bound to the galaxy, total gas enclosed by a sphere with the radius $\re$ and total gas within the disc, respectively. The last one is as before with the additional temperature threshold.
The mass of every black hole hosted by each galaxy is shown by a colour bar. We have chosen stellar mass as the indicator mass because it shows the tightest correlation with the magnetic field (see \cref{fig:Bd_Mgasre_Mstarre_cen}). The most striking feature in this figure is a clear break in the $\overline{B}-\Mstar$ increasing relation for $\Mstar \gtrsim 10^{10.3}$\msun in all four panels for star-forming galaxies with $\Mbh > 10^8$\msun (marked by an extra grey square around their points on the plot). This could be due to the fact that in the AGN feedback prescription of the TNG model, the low-accretion wind-driven kinetic feedback mode becomes dominant for black holes with $\Mbh \gtrsim 10^8$\msun \citep{Pillepich2021}. Restricting calculation of $\overline{B}$ to the gas cells within the effective radius and disc, results in a more scatter in comparison to the top panel. When we take the average over cold gas cells in the disc only, this scatter is decreased. More specifically, it seems there is a separate increasing trend for the most massive star-forming galaxies with $\Mstar \gtrsim 10^{11}$\msun. Moreover, it is also interesting that at the low mass end for $\bre$, galaxies that are still do not harbour any black hole represent stronger magnetic fields than those with the same stellar mass range, in average.

The treatment of the quenched galaxies (i.e. those with sSFR $<10^{-11}$yr$^{-1}$) depicted in the left panels are also of interest. One can see that the quenched galaxies with no active wind-driven AGN feedback ($\Mbh \lesssim 10^8$\msun) represent almost the same pattern in all the panels except the top one $(\btot)$. For other quenched galaxies with the turned on kinetic feedback mode ($\Mbh \gtrsim 10^8$\msun, denoted by an extra grey square), $\bre$, $\bd$ and $\bcgd$ show an increasing trend with $\Mstar$, if we ignore those with extremely low magnetic fields). As the figure obviously shows, the magnetic field of galaxies is strongly sensitive to the black hole driven feedback prescription. Therefore, we suggest it as a possible benchmark for validation of theoretical works.
%============
\begin{figure*}
	\centering
	\includegraphics[scale=0.4]{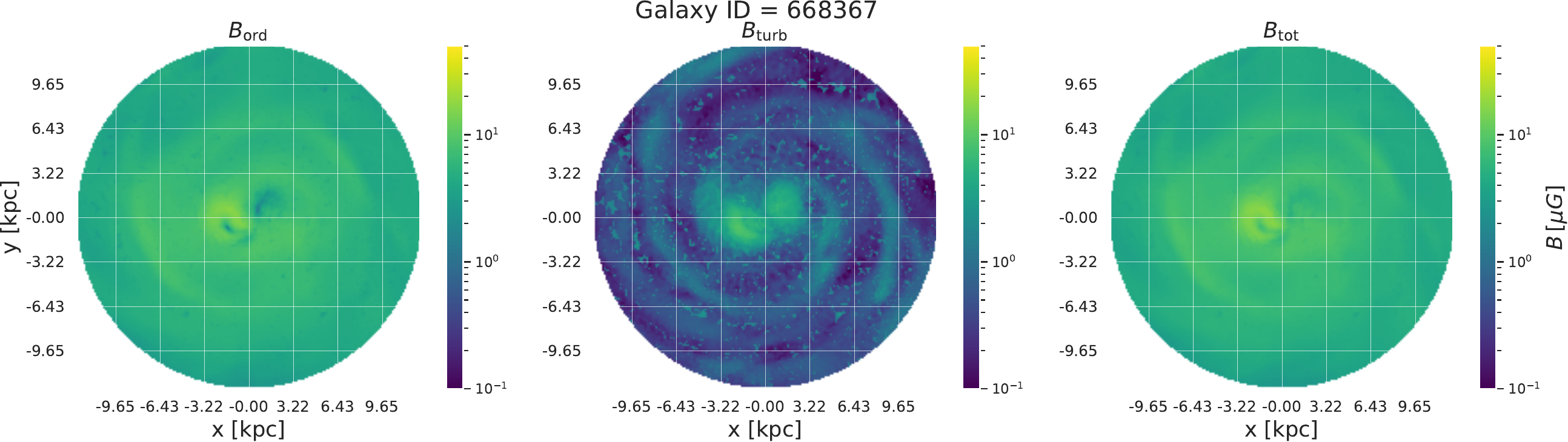}
	\centering
	\includegraphics[scale=0.4]{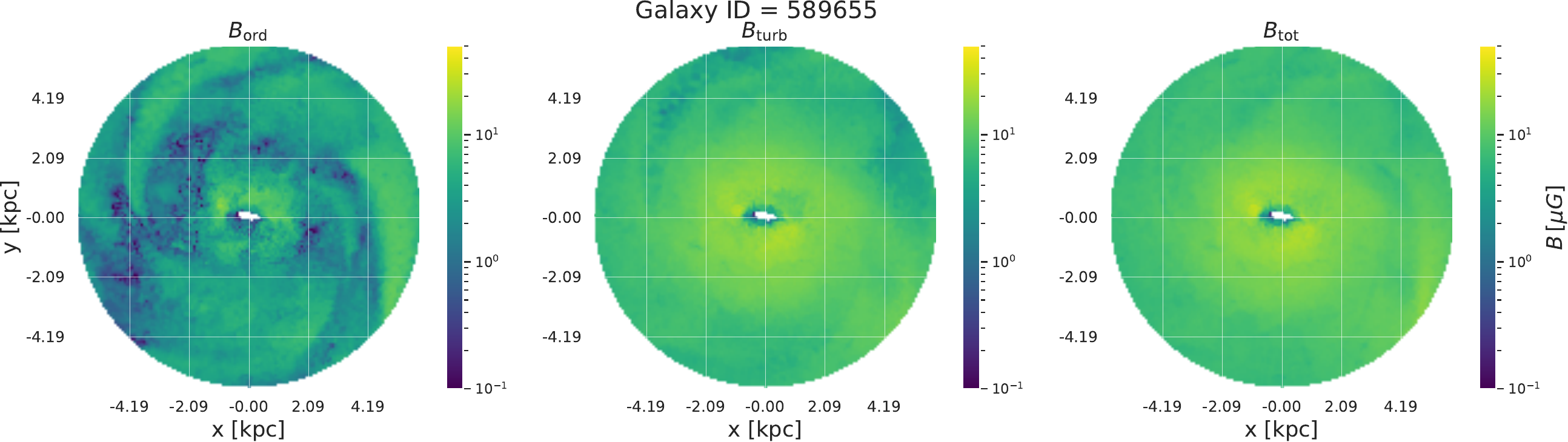}
	\caption{An example of the magnetic field in two galaxies with IDs = 668367 and 589655 (upper and lower row, respectively) within $\re$. From left to right, $B_{\rm ord}$, $B_{\rm turb}$ and $B_{\rm tot}$ are shown. In the upper case with $\btot\simeq 5.77\,\mu G$, the ordered component is dominant, $\bord/\bturb\simeq 4.79$ while in the lower case with $\btot\simeq 10.46\,\mu G$, the turbulent component is dominant, $\bord/\bturb\simeq 0.43$. Note that the color scales do not cover the complete range of $B$ to simplify comparison. The grid cell size in the upper and lower row is $\sim 64$ and 42 pc, respectively.}
	\label{fig:589655_re}
\end{figure*}
\begin{figure*}
	\centering
	\includegraphics[scale=0.45]{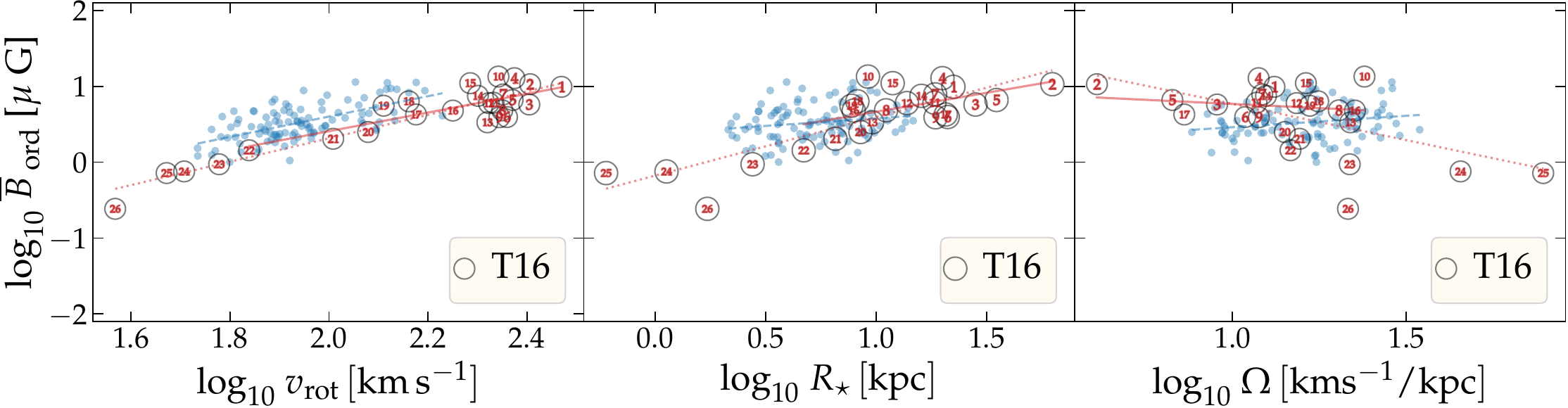}
	\includegraphics[scale=0.45]{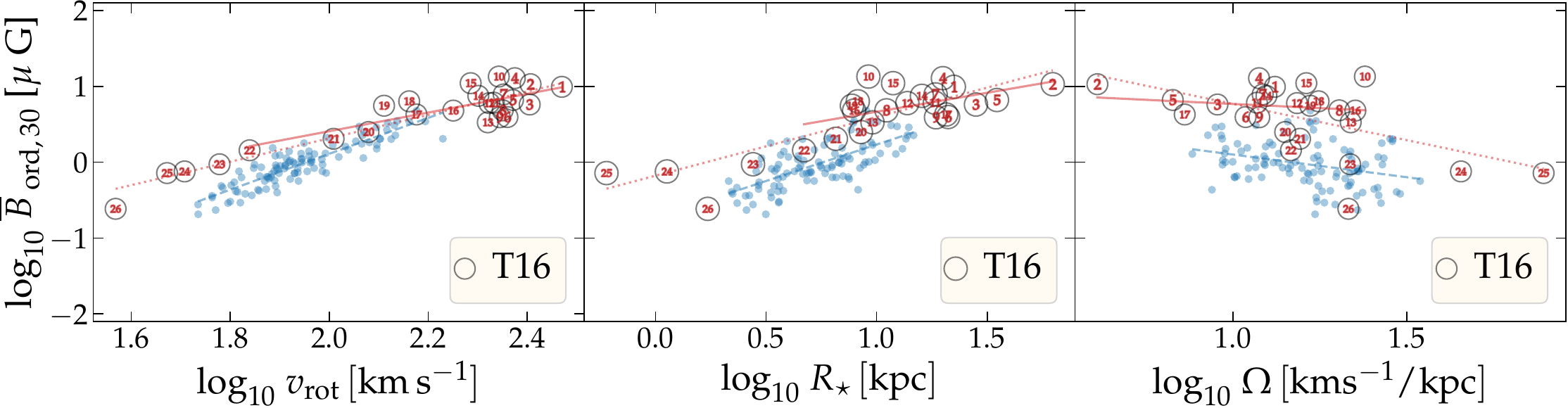}
	\caption{\textbf{Top}. $\bord$ vs. $\vrot$, $\re$ and $\Omega$. The blue circles show galaxies with $\Riso>2$ Mpc in the TNG50 sample. Overlaid circles denote the equipartition magnetic field vs. $\vrot$, $R_{\rm 25}$ and $\Omega$ as reported by \taba (See the text to see how they are compared). Numbers inside the circles are galaxy labels according to the first column of \cref{tab:gal}. The blue dashed and red dotted lines show the linear least-squares fit to the TNG50 and complete \taba galaxy samples, respectively whereas the red solid line shows this fit for \taba excluding galaxies \#: 23, 24, 25 and 26.
		\textbf{Bottom}. Same as the upper row, but here the perpendicular axis is $\bordth$ which is the ordered magnetic field calculated for a disc with the radius of 30 kpc.}
	\label{fig:Bd_Vrot_re_Omegare_cen_SF_3_ord_Riso2}
\end{figure*}
\begin{figure*}
	\centering
	\includegraphics[scale=0.45]{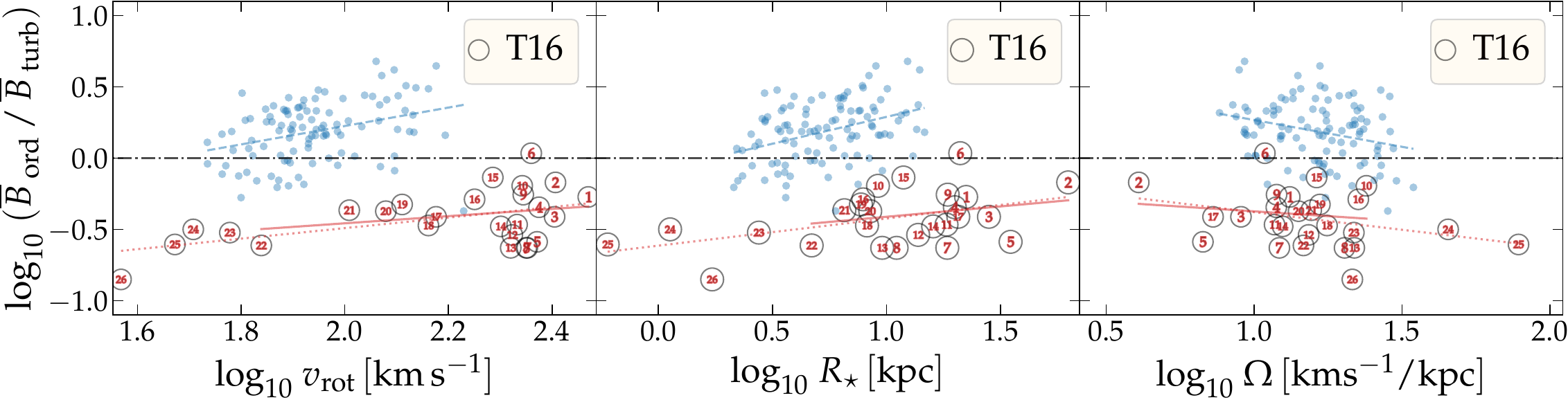}
	\caption{$\bord/\bturb$ vs. $\vrot$, $\re$ and $\Omega$. The horizontal black dot-dashed line denotes $\bord/\bturb=1$.}
	\label{fig:Bd_Vrot_re_Omegare_cen_SF_3_ordturb_Riso2}
\end{figure*}
\subsection{Comparison of TNG50 with observations}\label{sec:compare_with_T16}
In this section, our main goal is to compare the magnetic field of galaxies in the TNG50 sample with what \taba found for the ordered magnetic field of a sample of 26 nearby galaxies with already determined magnetic fields (see \cref{sec:empirical} for more details.). We also compare their turbulent and total magnetic fields with the simulation. Their sample (hereafter \taba sample) included 22 spirals plus 4 irregulars (taking LMC as irregular) .The \taba sample consisted of barred galaxies from \citet{Beck2002}, excluding those in the Ursa Major and Virgo clusters and the merging ones. They also considered dwarfs member of the Local Group as they also known to have large-scale regular magnetic field. We have listed all these galaxies in \cref{tab:gal}.
 
Regarding the TNG50 sample, to reduce the effect of environment and merging, we exclude galaxies with $\Riso \leqslant 2$ Mpc. This automatically removes the quenched galaxies (see \cref{sec:sf_or_q}) from our sample. Notice that most of the T16 galaxies are also star-forming or in transition to be quenched. This leaves us with a final sample of 103 galaxies. To better understand the effect of isolation and also an enhanced statistics, however, we repeat our analysis using the above criteria but with $\Riso > 1$ which will give 493 galaxies. As the more isolated galaxies are all star-forming, we exclude the only found three quenched galaxies from this sample. Notice that in this section, we do not distinguish between the cold and non-cold gas phases when calculating the average magnetic field, as this is also not done in \taba.

To best resemble the observation of radio synchrotron emission, firstly as before in \cref{sec:B_v_mass_corr_TNG}, we rotate all galaxies in our TNG sample, so the $z$ axis becomes the rotation axis. Then, for each galaxy using the \textsc{yt} package, we convert the unstructured Voronoi gas cells to a uniform 3D grid with the xy size of $r_{\rm{grid}}\times r_{\rm{grid}}$ and the width of 2 kpc that its centre coincide with the galaxy centre. We take $r_{\rm{grid}}$ as $\re$ or $r=30$ kpc which allows us to quantify the measurements for these two radii. Notice that, the aperture is an important factor for measuring the properties of galaxies. In addition, determining the size of galaxies is not a trivial task \citep[see e.g.,][]{Stevens2014a}. The cell size in these uniform grids is approximated as the typical size of the smallest Voronoi mesh in the galaxy. For each grid cell, we have $B_x$, $B_y$ and $B_z$ components. Using this approach, by averaging $B_x$ and $B_y$ along the $z$ axis in each cell of the grid, one can assume the ordered, turbulent and total magnetic field in the cells as
\begin{align}
B_{\rm ord\, i,j}  & = \sqrt{\langle B_{x\,\ijk}\rangle^2 + \langle B_{y\,\ijk}\rangle ^2}, \\
B_{\rm turb\, i,j}  & = \sqrt{\left\langle \left(B_{x\,\ijk} - \langle B_{x\,\ijk}\rangle \right)^2 + \left(B_{y\,\ijk} - \langle B_{y\,\ijk}\rangle \right)^2\right\rangle }, \\
B_{\rm tot\, i,j}  & = \sqrt{\langle B_{x\,\ijk}^2 + B_{y\,\ijk}^2\rangle }.
\end{align}
Here, i, j and k are the coordinate indices and angular bracket $\langle ... \rangle$ denotes averaging along the z axis. Therefore, for each cell in the grid we have
\begin{equation}
B^2_{\rm tot\,i,j}=B^2_{\rm ord\,\,i,\,j} + B^2_{\rm turb\,\,i,\,j}.
\end{equation}
Finally, we calculate $\btot$, $\bord$ and $\bturb$ by taking the average of corresponding $B$ values in each cell over a circle with the radius of $r_{\rm{grid}}$. In \cref{fig:589655_re} we have shown two gridded galaxies for instance from the TNG50 sample. The upper panels, show the magnetic fields components of a galaxy dominated by the ordered component whereas the lower panels show a galaxy with a dominant turbulent magnetic field.

\cref{fig:Bd_Vrot_re_Omegare_cen_SF_3_ord_Riso2} compares the TNG50 sample with $\Riso>2$ Mpc, against the \taba sample where we have plotted $\bord$ vs. $\vrot$, $\re$ and $\Omega$. In the \taba sample, $\vrot$ is the average of rotational velocity in the flat part of the rotation curve. Here, for the TNG50, we consider $\vrot$ as the average of $v_c$ (see \cref{eq:vc}) calculated at 100 points between the peak of the rotation curve, i.e. the maximum of $v_c$ and $r=100$ kpc and $\Omega=\vrot/\re$ (see \cref{sec:rotation_curve} for more details.).
On the other hand, \taba calculated the angular velocity as $\Omega=\vrot/R_{\rm{25}}$ with $R_{\rm{25}}$ the optical radius.
Considering equipartition between the energy density of the magnetic field and cosmic-rays which could be the case in the ISM of milky-Way-like galaxies, at scales $\gtrsim$ 1 kpc \citep{Seta2019a,Ponnada2022}, we calculate $\bord$, $\btot$ and $\bturb$ as suggested by \taba for their sample using the relations
\begin{align}
\btot &=\,\btoto \,\,\sqrt{\frac{S_{\rm I}}{S_{\rm I_0}} \, \left(\frac{\rm SFR_0}{{\rm SFR}}\right)^{0.5-0.6}}, \\
\bord &=\,\bordo \,\,\sqrt{\frac{S_{\rm PI}}{S_{\rm PI_0}} \, \left(\frac{\rm SFR_0}{{\rm SFR}}\right)^{0.5-0.6}}, \\
\bturb &= \sqrt{\btot^2 - \bord^2},
\end{align}
where, $S_{\rm I}$ and $S_{\rm PI}$ are the integrated flux densities of the total and linearly polarized intensities, respectively.
In these relations, quantities with 0 subscripts are values of a reference galaxy taken to be M33 (\# 21 in the \cref{tab:gal} and \cref{fig:Bd_Vrot_re_Omegare_cen_SF_3_ord_Riso2}) and are $\btoto = 6.4\, \rm{\mu G}$, $\bordo=2.5\,\rm{\mu G}$ \citep{Tabatabaei2008a}, $S_{\rm I_0}=9.06\,\rm{mJy}$, $S_{\rm PI_0}=0.53\,\rm{mJy}$ and $\rm{SFR_0}=0.3\rm{M_{\odot}\,yr^{-1}}$. We have listed $\Pearson$, $\Spearman$ and the slope of a least-squares fit $a$ calculated for the TNG50 and \taba samples in \cref{tab:TNG_vs_T16} as well as a sub-sample of the latter namely T16\_22 in which the irregular galaxies \#: 23, 24, 25 and 26 (SMC, NGC 6822, IC 10 and IC 1613) are excluded.

In the upper row of \cref{fig:Bd_Vrot_re_Omegare_cen_SF_3_ord_Riso2}, the left, middle and right panels show $\bord$ vs. $\vrot$, $\re$ and $\Omega$, respectively. A glance at the upper-right panel, reveals an excellent agreement between the TNG50 and \taba sample. In the former, $a=1.34$ with the Pearson and Spearman rank coefficients $\Pearson=0.63$ and $\Spearman=0.57$ whereas in the latter $a=1.54$ with tighter correlations of $\Pearson=0.92$ and $\Spearman=0.73$. The sample T16\_22 even demonstrates a better agreement with $a=1.23$. As we have restricted the TNG50 sample to the galaxies with $
\Riso>2$ Mpc, galaxies with $\vrot \gtrsim 160\,\kms$ are not included. Including galaxies with $\Riso>1$ Mpc, which we will discuss its effects in \cref{sec:B_corr_with_Riso1_T26}, increases this limit to $\sim 230\,\kms$.  We next consider $\bord$ vs. $\re$ in the middle panel. Here, the correlation coefficients are $\Pearson=0.82$ and $\Spearman=0.65$ in the \taba sample. More specifically, ignoring the smallest galaxies, yields weaker correlations ($\Pearson=0.53$ and $\Spearman=0.43$) and also a shallower slope, thus a better compromise with the TNG50 that shows very weak correlations of $\Pearson=0.18$ and $\Spearman=0.17$, with $a=0.20$.
In the right panel, the correlation with $\Omega$ is examined.
%In addition, we examine the correlation of $\bord$ with $\Omega$.
It can be observed that while the \taba sample shows a more or less inverse correlation with $\Pearson=-0.55$, the TNG50 sample has $\Pearson=0.2$. The T16\_22 is in better agreement with the simulation with $\Pearson=-0.16$.
Furthermore, \taba found no correlation between $\bord$ and $\vrot/R_{\rm{max}}$ with $R_{\rm{max}}$ the radius where $v_c$ peaks.

As already mentioned, we repeat the previous analysis for $\bordth$, the average magnetic field in all the bound gas cells within a disc with the same width (1 kpc) but with a fix cut-off radius of 30 kpc. The lower row in \cref{fig:Bd_Vrot_re_Omegare_cen_SF_3_ord_Riso2} compares $\bordth$ for the TNG50 sample against the \taba sample. Obviously, the average magnetic field in 30 kpc has decreased for all the TNG50 galaxies with respect to the average in $\re$. This decrease which is stronger for the smaller galaxies due to a weaker magnetic field pervading them, yields a steeper trend ($a=2.37$) when $\bordth$ is plotted vs. $\vre$. Moreover, the correlation becomes tighter ($\Pearson=0.9$). More interestingly, the correlation between $\bordth$ and $\re$ has also become tighter ($\Pearson=0.7$) and the slope has increased to 0.96. This is in contrast to the no correlation of $\bord$ vs. $\re$ as can be seen from the middle panel in the the upper row. It is also remarkable that now the trend of $\bordth-\Omega$ has a negative slope $a=-0.6$ with a Pearson coefficient $\Pearson=-0.32$.

In \cref{tab:TNG_vs_T16}, in addition to $\bord$, we have also listed $\Pearson$, $\Spearman$ and $a$ for $\bturb$ and $\btot$. From the table, we can see that same as the \taba sample, the correlation coefficients and the slope are decreased from $\bord$ to $\bturb$ for $\vrot$. Moreover, the values show that the \taba sample is more correlated ($\Pearson=0.83$) than the TNG50 one ($\Pearson=0.41$), though the $\bturbth-\vrot$ correlation is stronger ($\Pearson=0.93$). For $\re$, while the \taba and T16\_22 samples both have positive correlations ($\Pearson=0.73$ and 0.36, respectively), the TNG50 galaxies represent an inverse weak correlation ($\Pearson=-0.20$), in spite the fact that $\bturbth$ and $\re$ show positive connection. This in turn makes $\bturbre-\Omega$ correlation to significantly differ in the observation and simulation ($\Pearson=-0.49$, -0.06 and 0.57, in the T16, T16\_22 and TNG50, respectively). However, $\bturbth$ with $\Omega$ shows a weak relation ($\Pearson=-0.29$), i.e. in better agreement with observations. Finally, as one would expect, the least-squares fitted slopes for $\btot$ must be between those values for $\bord$ and $\bturb$ . We can also see that for the \taba sample, the two correlations $\btot-\vre$ and $\btot-\re$ are a little weaker than $\bord-\vre$ and $\bord-\re$. This is also the case for $\btot-\re$ in our TNG50 sample, but not for $\btot-\vre$. For $\btotth$ both of the trends are very slightly stronger than $\bordth$.
%============
\begin{table}
	\centering
	\begin{threeparttable}  
		\caption{Properties of the galaxy sample in \citetalias{Tabatabaei2016}.}
		\label{tab:gal}
		\begin{tabular}{ccccc}
			\hline
			%$ $ & $ $ &\multicolumn{2}{c}{$\kappa$} &\multicolumn{2}{c}{$n$} &\multicolumn{1}{c}{$A$}\\ 
			\# &  Galaxy Name &Hubble Type & $\dfrac{v_{\rm rot}}{\rm{(km/s)}}$ & $\dfrac{R_{25}}{\rm{(kpc)}}$ \\ 
			\hline
			1 &	 NGC 1097  			&     	SBbc  & 295\,$\pm$\,24  & 22.35\\
			2 &	 NGC 4565  			&  		SAb    &   255\,$\pm$\,12   & 62.61\\ 
			3 &	 NGC 5907  			&  		Sc   &  254\,$\pm$\,14  & 28.14\\ 
			4 &	 NGC 7479 			&   	SBbc & 237\,$\pm$\,18    & 19.94\\ 
			5 &	 NGC 1365  			&     	SBb &  235\,$\pm$\,15   & 35.01\\ 
			6 &	 M31     			&   	SAb   &   229\,$\pm$\,12 & 21.06\\
			7 &	 NGC 891   			&  		SAb    & 225\,$\pm$\,10   & 18.48\\
			8 &	 NGC 7552 			&  		SBbc &  224\,$\pm$\,29   & 11.08\\
			9 &	 NGC 1300 			&  		SBb &  221\,$\pm$\,20   & 18.56\\
			10 & NGC 6946 			&   	SABcd  &  220\,$\pm$\,24    & 9.18\\
			11 & NGC 3628  			&  		SAb\,pec &  215\,$\pm$\,15  & 18.34\\
			12 & NGC 253			&    	SABc        &  211\,$\pm$\,12  & 13.77\\
			13 & NGC 5643 			&  		SBc  &  209\,$\pm$\,27  &  9.58\\
			14 & NGC 1672  			&   	SBb & 200\,$\pm$\,15  & 16.02\\
			15 & IC 342   			&    	SABcd   &   193\,$\pm$\,24   & 11.87\\
			16 & NGC 4736 			&   	SABab  &  178\,$\pm$\,16   & 7.91\\  
			17 & NGC 3359  			&    	SBc &  149\,$\pm$\,6  & 20.65\\ 
			18 & NGC 1559  			&   	SBc &  145\,$\pm$\,9   & 8.22 \\
			19 & NGC 3059 			&   	SBc & 129\,$\pm$\,16   &  7.73\\
			20 & M33     			&    	SAcd &  120\,$\pm$\,10  & 8.49\\
			21 & NGC 1493 			&    	SBc &  102\,$\pm$\,12   &  6.54\\
			22 & LMC     			&  		Irr/SBm  &  69\,$\pm$\,7 & 4.70\\
			23 & SMC     			&    	Irr   &  59\,$\pm$\,4.5  & 2.75\\ 
			24 & NGC 6822   		&     	Irr &   51\,$\pm$\,4  & 1.13\\
			25 & IC 10    			&    	Irr &  47\,$\pm$\,5  & 0.60\\ 
			26 & IC 1613  			&     	Irr &   37\,$\pm$\,5  & 1.72\\
%			1 &	 IC342   &    SABcd   &   193\,$\pm$\,24   & 11.87\\
%			2 &	 NGC6946 &   SABcd  &  220\,$\pm$\,24    & 9.18\\
%			3 &	 NGC253  &    SABc        &  211\,$\pm$\,12  & 13.77\\
%			4 &	 NGC3628  &  SAb\,pec &  215\,$\pm$\,15  & 18.34\\
%			5 &	 NGC4565  &  SAb    &   255\,$\pm$\,12   & 62.61\\
%			6 &	 NGC4736 &   SABab  &  178\,$\pm$\,16   & 7.91\\  
%			7 &	 NGC5907  &  Sc   &  254\,$\pm$\,14  & 28.14\\
%			8 &	 NGC891   &  SAb    & 225\,$\pm$\,10   & 18.48\\
%			9 &	 NGC1097   &     SBbc  & 295\,$\pm$\,24  & 22.35\\  
%			10 &	 NGC1365  &     SBb &  235\,$\pm$\,15   & 35.01\\ 
%			11 &	 NGC3359  &    SBc &  149\,$\pm$\,6  & 20.65\\ 
%			12 &	 NGC1493 &    SBc &  102\,$\pm$\,12   &  6.54\\
%			13 &	 NGC1559  &   SBc &  145\,$\pm$\,9   & 8.22 \\
%			14 &	 NGC1672  &   SBb & 200\,$\pm$\,15  & 16.02\\
%			15 &	 NGC3059 &   SBc & 129\,$\pm$\,16   &  7.73\\
%			16 &	 NGC5643 &  SBc  &  209\,$\pm$\,27  &  9.58\\
%			17 &	 NGC7552 &  SBbc &  224\,$\pm$\,29   & 11.08\\
%			18 &	 NGC1300 &  SBb &  221\,$\pm$\,20   & 18.56\\
%			19 &	 NGC7479 &   SBbc & 237\,$\pm$\,18    & 19.94\\ 
%			20 &	 M31     &   SAb   &   229\,$\pm$\,12 & 21.06\\
%			21 &	 M33     &    SAcd &  120\,$\pm$\,10  & 8.49\\
%			22 &	 LMC     &  Irr/SBm  &  69\,$\pm$\,7 & 4.70\\
%			23 &	 SMC     &    Irr   &  59\,$\pm$\,4.5  & 2.75\\ 
%			24 &	 IC10    &    Irr &  47\,$\pm$\,5  & 0.60\\ 
%			25 &	 NGC6822   &     Irr &   51\,$\pm$\,4  & 1.13\\
%			26 &	 IC1613  &     Irr &   37\,$\pm$\,5  & 1.72\\
			\hline
		\end{tabular}
		\begin{tablenotes}  
			\item[a] \textit{Note}. Galaxies are sorted according to their rotational velocity $v_{\rm{rot}}$. For other galaxy properties, see \citetalias{Tabatabaei2016} where these date are taken from.
		\end{tablenotes}  
	\end{threeparttable}
\end{table}

\begin{table*}
	\centering
		\caption{Correlation of the ordered, turbulent and total magnetic fields, for the TNG50 sample galaxies with $\Riso>2$ Mpc and for the T16 samples with $\vrot$, $\re$ and $\Omega$. See the text for the definition of $\vrot$ and $\Omega$. Magnetic field components for the TNG50 galaxies are calculated both for discs of radii $\re$ and 30 kpc. The latter denoted by a subscript 30. T16\_22 is the T16 sample without irregular galaxies.}
		\label{tab:TNG_vs_T16}
		\begin{tabular}{ccccccccccc}
			\hline
			%$ $ & $ $ &\multicolumn{2}{c}{$\kappa$} &\multicolumn{2}{c}{$n$} &\multicolumn{1}{c}{$A$}\\ 
			Sample & Property & $\Pearson$ & $\Spearman$ & $a$ & Sample & $\Pearson$ & $\Spearman$ & $a$ \\
			\hline
%% TNG ========================================================================
			TNG50,\,$\bord$      & $\vrot$       & 0.63 & 0.57 & 1.34 & TNG50,\,$\bordth$ & 0.90 & 0.88 & 2.37 \\
			..                   & $\re$         & 0.18 & 0.17 & 0.20 & .. & 0.70 & 0.70 & 0.96 \\
			..                   & $\Omega$    & 0.20 & 0.18 & 0.30 & .. & -0.32 & -0.34 & -0.60 \\
			                     \\
			TNG50,\,$\bturb$     & $\vrot$       & 0.41  & 0.30  & 0.70  & TNG50,\,$\bturbth$ & 0.93 & 0.92 & 1.89 \\	
			..                   & $\re$         & -0.20 & -0.22 & -0.18 & .. & 0.70 & 0.74 & 0.74 \\		
			..                   & $\Omega$    & 0.57  & 0.54  & 0.68  & .. & -0.29 & -0.37 & -0.42 \\
			                     \\
			TNG50,\,$\btot$      & $\vrot$       & 0.65 & 0.55 & 1.18 & TNG50,\,$\btotth$ & 0.93 & 0.91 & 2.28 \\	
			..                   & $\re$         & 0.08 & 0.06 & 0.08 & .. & 0.71 & 0.71 & 0.90 \\
			..                   & $\Omega$    & 0.34 & 0.32 & 0.45 & .. & -0.31 & -0.35 & -0.54 \\
								 \\
		TNG50,\,$\bord/\bturb$	 & $\vrot$       & 0.34 & 0.35 & 0.65     & TNG50,\,$\bordth/\bturbth$ & 0.33 & 0.32 & 0.48 \\
			..                   & $\re$         & 0.38 & 0.35 & 0.38     & .. & 0.30 & 0.29 & 0.22 \\
			..                   & $\Omega$    & -0.28 & -0.24 & -0.38  & .. & -0.17 & -0.18 & -0.18 \\
%% T16 ========================================================================
			                     &               &   & & & \\			                     
			T16,\,$\bord$		 & $\vrot$       & 0.92  & 0.73  & 1.54  & T16\_22,\,$\bord$ & 0.74 & 0.56 & 1.23 \\
			..                   & $\re$         & 0.82  & 0.65  & 0.77  & .. & 0.53 & 0.43 & 0.50 \\
		    ..                   & $\Omega$    & -0.55 & -0.40 & -0.96 & .. & -0.16 & -0.11 & -0.22 \\
			                     &               &   & & & \\
			T16,\,$\bturb$		 & $\vrot$       &  0.83  & 0.68  & 1.17  & T16\_22,\,$\bturb$ & 0.56 & 0.48 & 0.98 \\
			..                   & $\re$         &  0.73  & 0.59  & 0.58  & .. & 0.36 & 0.32 & 0.36 \\
			..                   & $\Omega$    &  -0.49 & -0.37 & -0.71 & .. & -0.06 & -0.05 & -0.08 \\
			&               &   & & & \\
			T16,\,$\btot$		 & $\vrot$       &  0.87  & 0.71  & 1.22  & T16\_22,\,$\btot$ & 0.63 & 0.53 & 1.04 \\
			..                   & $\re$         &  0.77  & 0.61  & 0.61  & .. & 0.41 & 0.36 & 0.39 \\
			..                   & $\Omega$    &  -0.51 & -0.37 & -0.75 & .. & -0.08 & -0.06 & -0.11 \\
								\\
	T16,\,$\bord/\bturb$		 & $\vrot$       & 0.49 & 0.37 & 0.37  & T16\_22,\,$\bord/\bturb$ & 0.21 & 0.18 & 0.25 \\
			..                   & $\re$         &  0.45 & 0.38 & 0.19  & .. & 0.21 & 0.18 & 0.15 \\
			..                   & $\Omega$    &  -0.32 & -0.28 & -0.25 & .. & -0.14 & -0.10 & -0.13 \\
			\hline
		\end{tabular}
\end{table*}

%============
\subsection{Comparison of the ordered-to-turbulent magnetic field ratio}\label{sec:ord_to_turb_ratio}
\cref{fig:Bd_Vrot_re_Omegare_cen_SF_3_ordturb_Riso2} demonstrates $\bord/\bturb$ ratio for the TNG50 and \taba samples. It shows that for the most of galaxies in the TNG50 this ratio is greater than one, while for the \taba sample is smaller. Only the galaxy \#6, the M31 has a little greater value. Moreover, this ratio is increasing with respect to $\vrot$ and $\re$ and decreasing with regard to $\Omega$. We should notice that the maximum resolution element of TNG50 for the gaseous matter is $\sim 100$ pc which is not adequate to resolve the most of the turbulent magnetic fields that are living in smaller scales. More specifically, owing to the incorporation of an effective EOS in the TNG50 (and to our knowledge, all the carried out large-scale galaxy formation simulation until now) as a proxy for the unresolved ISM \citep{Springel2003}, the distorted flows caused by supernovae ejections are ignored. 

Note that in the TNG model, galactic outflows are produced by supernova-driven wind particles \citep{Pillepich2018a}. These particles are hydrodynamically decoupled from their surrounding high density gas, until they inter the low background medium generally within a few kpcs, when they deposit their thermal energy, momentum, mass and metal into the gas cells in which they are located. This means the supernova induced small-scale dynamo is overlooked in the TNG model, although the turbulent motions are captured indirectly by processes like galactic fountains. Moreover, the central AGN feedback is also capable to directly affect the cold and dense ISM. 

It should also be pointed out that in observations, the beam depolarization could decrease the strength of ordered field, leading to an overestimated turbulent component \citep{Sokoloff1998}.
%============
\subsection{Numerical MHD considerations}\label{sec:mhd_problems}
The lack of sufficient resolution still inhibits proper modeling of the magnetic field in all types of the MHD simulations of galaxy evolution. The magnetic Prandtl (the ratio of kinematic viscosity and magnetic diffusivity) and Reynolds numbers in the ISM tend to be very large \citep{Ferri2020}, orders of magnitude larger than values in the current MHD simulations \citep{Federrath2014b}.

Usage of the ideal MHD in the cosmological MHD simulations (including the TNG suite) is currently the only doable option, as incorporating the non-ideal terms requires much more computation time. However, it is also likely that adding viscosity and resistivity at their physical values, do not change anything at the current available resolution of cosmological simulations. The ideal MHD approximation is equivalent to an effective magnetic Prandtl number $\sim1$ as dissipation due to the fluid viscosity and magnetic resistivity is not realistically implemented.

For the Prandtl number near the unity, the magnetic field can be amplified by turbulence induced dynamo, when the Reynolds number exceeds the critical value of $30-35$ \citep[e.g.,][]{Brandenburg2005a}. One can estimate the effective Reynolds number as $(L/\epsilon\Delta x)^{4/3}$, where $L$ is the typical length scale or the turbulent outer scale\footnote{Also called the injection scale.} of the system, $\Delta x$ is the resolution element and $\epsilon$ is a factor determined by the diffusivity of the numerical method which is of order unity for second order finite volume codes like \textsc{arepo} \citep{Donnert2018,McKee2020a}. As a rough estimate, for $L\sim 1$ kpc (the typical size of the thick disc), if it has been resolved entirely by the maximum resolution of $\Delta x\sim 100$ pc, the effective Reynolds number will be $\sim 22$. This is smaller than the critical value proposed to set off the turbulent dynamo. Thus magnetic fields should have been amplified mostly by gas condensation or to some extent by shear caused by the galactic differential rotation which in turn could generate ordered fields that are intrinsically anisotropic turbulent fields \citep{Beck2019}.

The insufficient resolution of MHD simulations also prevents these models to reach to the observed energy equipartition between turbulence and magnetic field \citep{Crutcher2009,Basu2013,Lopez-Rodriguez2021}. This could lead to the weaker turbulent magnetic fields as can be seen in \cref{fig:Bd_Vrot_re_Omegare_cen_SF_3_ordturb_Riso2}. Moreover, the lack of enough resolution could slow down and finally stop the amplification of magnetic field by small-scale dynamo (see \cref{sec:effect_of_low_res} for a comparison of TNG50-1 with its lower-resolution box).

In addition, another issue which stems from the difficulty in maintaining the solenoidal constraint $\nabla\cdot{B}=0$, adds to the challenge of modeling of MHD. As we already mentioned in \cref{sec:B_in_TNG}, the TNG simulation suit has been run with a version of the \textsc{arepo} code that applies the eight-wave Powell cleaning method for this purpose, but it suffers from unwanted numerical side-effects. Another approach that ensures the divergence-free condition to machine precision is the constraint transport \citep{Evans1988}. \cite{Mocz2016} successfully implemented this method in the \textsc{arepo} code. Comparing with the divergence cleaning scheme, they found that in the complex astrophysical flows, the constraint transport is generally gives more reasonable results. For instance, they observed that in a turbulent medium, the cleaning method gives rise to a spurious growth of the magnetic field. More importantly, this also happens in simulation of an idealized isolated disc galaxy, specially in its centre where the divergence error could be as large as 10 percent of the total gas pressure. Moreover, the morphology of the magnetic field patterns is also affected by the divergence error. Whereas the constraint transport shows the winding of the magnetic fields at early stages of simulation, the Powell scheme does not (see their figure 7).

Despite the above arguments, the results of several works done by \textsc{arepo} that utilized the divergence cleaning scheme, argued to be in agreement with the observed properties of galaxies \citep{Pakmor2017,Pakmor2018,Pakmor2020}.It should also be noted that, using an adaptive-mesh refinement code equipped with the constraint transport method as well as a subgrid model for the mean field dynamo, \citet{Liu2022} performed a zoom-in simulation of the magnetic field evolution in a Milky way-like galaxy. They found a magnetic field structure similar to what previously generated in the Auriga zoomed-in galaxies \citep{Pakmor2020} although with much powerful primordial seeds. As a result, we think this still needs more detailed studies.
%============
\section{Summary}\label{sec:summary}
Using the high-resolution TNG50 large cosmological galaxy formation simulation, we investigate the correlation of the magnetic field with the gaseous, stellar and total mass of central disc galaxies with $\Mstar>10^8$\msun. For this sample of galaxies which we call it the TNG50 sample, we also study correlations with other galaxy properties such as SFR, $D/T$ and $\mathcal{F}$. In addition, we consider the effect of relative isolation, gas temperature and AGN feedback on the correlations. We further compare the simulation with the \taba finding. Our main results are as follows:
\begin{itemize}
	\item A correlation holds between the magnetic field pervaded the disc of isolated galaxies with $\Riso>2$ Mpc and their total mass in $\re$ as $\bd\sim\Mtotre^{0.19}$ ($\Pearson=0.41$ and $\Spearman=0.37$) in the TNG50 sample.
	\item For this isolated sample, we also find a tight correlation ($\Pearson=0.64$ and $\Spearman=0.58$) with circular velocity at $\re$ as $\bd\sim\vre^{1.01}$.
	\item The above relations also hold considering $\bcgd$ ($\bbar$ calculated in the cold gas in the disc) with $T<5\times10^4\kelvin$.
	\item Galaxies transiting into quiescence (sSFR $\approx10^{-11}\rm{yr}^{-1}$) or those already quenched (sSFR $<10^{-11}\rm{yr}^{-1}$) do not follow the above trends.
	\item Galaxies with equal $\Mgre$ and lower gas fraction, have stronger $\bd$. This could be indicative of gas flow to the CGM by magnetic field-driven outflows and wind.
	\item The stellar mass of galaxies is more correlated with their magnetic fields, compared to the gaseous and the total masses.
	\item The wind-driven black hole feedback implementation in the TNG galaxy formation model, causes a break in the correlations mentioned above at $\Mstar\gtrsim 10^{10.3}$. 
	Hence, the similar deviation found in real observations (\taba) can be linked to the AGN feedback. 
	\item Assuming equipartition between the energy density of the magnetic field and cosmic-rays, \taba found a tight correlation ($\Pearson=0.92$) between $\bord$ and $\vrot$ in a sample of non-cluster non-interacting nearby galaxies as $\bord\sim\vrot^{1.54}$. In agreement with their result, we find that for the TNG50 star-forming sample, $\bord$ is correlated with $\vrot$ ($\Pearson=0.63$), with a very similar slope ($a=1.34$).
	\item In the \taba sample, $\bord$ is only weakly correlated with $R_{25}$. We find no correlation between $\bord$ and $\Omega$ based on TNG50 which agrees with observations by T16.
	\item The $\bord/\bturb$ ratio is larger than one for the TNG50 star-forming sample, but in \taba, the ratio is smaller than one. This could be due to the fact that the synchrotron polarization observations suffer from depolarization effects reducing the ordered field compared to the turbulent component. It is also possible that the turbulent component is not resolved in the TNG50 simulation. 
	\item The correlation of the magnetic field with galaxy properties is found to be tighter in the more isolated TNG50 samples. 
	\item The magnetic field correlations are tighter in the colder than warmer gas phase for the TNG50 samples with an equal degree of isolation.
\end{itemize}
As a final remark, it is necessary to mention that although the magnetic field distribution in TNG50 galaxies seems to be in agreement with the current observations, there are some serious tensions between TNG50 and other galactic scale observations. To be specific, it is shown in \citet{Roshan2021a} that the stellar bars in TNG50 are ``slow" because of the secular interaction with the dark matter halo. The dynamical friction caused by the halo particles slows down the pattern speed of the bars. On the other hand, the majority of the observed bars are fast. Recently, another tension has been reported in \citet{Kashfi}. It turns out that the fraction of barred submaximal discs in TNG50 is significantly smaller than in SPARC data set. This is related to the stabilizing impact of the dark matter halo on suppressing the bar instability. Furthermore, TNG50, in gross disagreement with observation, is unable to reproduce long stellar bars \citep{Frankel} and thin galactic discs \citep{Moritz}. Anyway, it seems that standard cosmological simulations are still far away from reproducing all the observed properties of the galaxies. Therefore huge numerical effort and also theoretical investigations on the nature of dark matter particles are still required to reconcile cosmological simulated galaxies with real observations.
%================================
\section*{Acknowledgements}
The authors thank the anonymous referee for insightful comments on the paper draft. They also thank Rainer Beck for valuable comments. MH thanks Najme Mohammad-Salehi, Golshan Ejlali, Mohammad Javad Shahhoseini, Tahere Kashfi, Hamid Hassani, Masoumeh Ghasemi-Nodehi, Mohammad Reza Nasirzadeh and Maryam Khademi for useful discussions.
This research made use of \texttt{NUMPY}~\citep{numpy}, \texttt{SCIPY}~\citep{scipy}, \texttt{ASTROPY}~\citep{astropy}, \texttt{H5PY}~\citep{h5py}, \texttt{SCIKIT-LEARN}~\citep{scikit-learn} and \texttt{JUPYTER}~\citep{jupyter}. All figures were generated using \texttt{MATPLOTLIB}~\citep{matplotlib}. We used \texttt{SEABORN}~\citep{seaborn} and \texttt{YT}~\citep{yt} to plot \cref{fig:kde_density_plot_B} and \cref{fig:galaxies}. The main part of the calculation was performed at the \href{https://scicluster.readthedocs.io}{Sci-HPC} center of the Ferdowsi University of Mashhad. Also, we have made extensive use of the NASA Astrophysical Data System Abstract Service.
%%%%%%%%%%%%%%%%%%%%%%%%%%%%%%%%%%%%%%%%%%%%%%%%%%
\section*{Data availability}
This project is based on the publicly available \href{https://www.tng-project.org}{TNG-project} database. The reduced data used in this contribution are available upon request.
\bibliographystyle{mnras}
\bibliography{library} % if your bibtex file is called example.bib
%%%%%%%%%%%%%%%%% APPENDICES %%%%%%%%%%%%%%%%%%%%%
\begin{table*}
	\centering
		\caption{Correlation of the ordered, turbulent and total magnetic fields, for the TNG50 sample galaxies with $\Riso>1$ Mpc and for the T16 samples with $\vrot$, $\re$ and $\Omega$. See the text for the definition of $\vrot$ and $\Omega$. Magnetic field components for the TNG50 galaxies are calculated both for discs of radii $\re$ and 30 kpc. The latter denoted by a subscript 30. T16\_22 is the T16 sample without irregular galaxies.}
		\label{tab:TNG_vs_T16_Riso1}
		\begin{tabular}{ccccccccccc}
			\hline
			%$ $ & $ $ &\multicolumn{2}{c}{$\kappa$} &\multicolumn{2}{c}{$n$} &\multicolumn{1}{c}{$A$}\\ 
			Sample & Property & $\Pearson$ & $\Spearman$ & $a$ & Sample & $\Pearson$ & $\Spearman$ & $a$ \\
			\hline
%% TNG ========================================================================
			TNG50,\,$\bord$      & $\vrot$       & 0.46  & 0.43  & 0.95  & TNG50,\,$\bordth$ & 0.83 & 0.83 & 1.95 \\
			..                   & $\re$         & -0.09 & -0.09 & -0.10 & .. & 0.62 & 0.62 & 0.76 \\
			..                   & $\Omegare$    & 0.41  & 0.35  & 0.53  & .. & -0.24 & -0.28 & -0.36 \\
			                     \\
			TNG50,\,$\bturb$     & $\vrot$       & 0.39  & 0.33  & 0.79  & TNG50,\,$\bturbth$ & 0.90 & 0.89 & 1.76 \\
			..                   & $\re$         & -0.36 & -0.34 & -0.37 & .. & 0.54 & 0.55 & 0.55 \\
			..                   & $\Omegare$    & 0.69  & 0.63  & 0.88  & .. & -0.10 & -0.18 & -0.12 \\
			                     \\
			TNG50,\,$\btot$      & $\vrot$       & 0.51  & 0.47  & 0.94  & TNG50,\,$\btotth$ & 0.88 & 0.88 & 1.93 \\
			..                   & $\re$         & -0.19 & -0.18 & -0.19 & .. & 0.62 & 0.62 & 0.71 \\
			..                   & $\Omegare$    & 0.56  & 0.50  & 0.66  & .. & -0.21 & -0.25 & -0.29 \\
								 \\
		TNG50,\,$\bord/\bturb$	 & $\vrot$       & 0.08 & 0.12 & 0.16     & TNG50,\,$\bordth/\bturbth$ & 0.13 & 0.15 & 0.20 \\
			..                   & $\re$         & 0.27 & 0.25 & 0.27     & .. & 0.26 & 0.27 & 0.21 \\
			..                   & $\Omegare$    & -0.28 & -0.24 & -0.34  & .. & -0.24 & -0.23 & -0.24 \\
			\hline
		\end{tabular}
\end{table*}

\appendix
\section{The empirical relation}\label{sec:empirical}
\citetalias{Tabatabaei2016} investigated the existence of a relation between the average large-scale ordered magnetic field $\bord$\footnote{Notice they used $\overline{B}$ symbol for this kind of magnetic field.} and the rotational velocity of galaxies. For this purpose, they used radio synchrotron polarization and rotational velocity data of spiral/irregular non-interacting field galaxies available in the literature. Their sample included 26 nearby galaxies from which 22 were spirals (see \cref{tab:gal}). The rotational velocities were determined as the mean value of the flat part of the rotation curves, or from the 20 per cent of the width of HI line profiles $(W_{20})$ from HIPASS catalog or from the literature. 

To estimate the total and ordered (large-scale) magnetic field,  the total $(\SI)$ and polarized $(\SPI)$ radio emissions at 4.8 GHz were used respectively. We know that the ratio of polarized intensity to number of cosmic-rays $\SPI/\Ncr$ is proportional to $\bord$. Therefore, the estimation of $\bord$ could be done via two approaches: first, taking the SFR as a proxy for $\Ncr$ simply yields: $\bord \sim \SPI/\rm {SFR}$. Second, assuming an equipartition between magnetic field and cosmic-ray energy densities which leads to $\Ncr \sim B^2$ and then using $B\sim\rm {SFR}^{0.25-0.3}$ relation from independent observations \citep[e.g.][]{Heesen2014}, one can obtain $\bord \sim \SPI/\rm {SFR}^{0.5-0.6}$. To estimate SFR, they followed the calibration relations provided by \citet{Hao2011} and \citet{Kennicutt2012}.

Using the first approximation, they found $\bord \sim v_{\rm rot}$ whereas $\bord \sim v_{\rm rot}^{1.5}$ for the second one. More specifically, they obtained such a relation for the dynamical mass of their sample of galaxies $M_{\rm {dyn}}\simeq R_{\rm {25}} v_{\rm {rot}}^2/G$ where $R_{\rm {25}}$ is the mass inside the optical radius and $G$ is the gravitational constant. For the two approximations, they correspondingly found $\bord \sim M_{\rm dyn}^{\,0.25}$ and $\bord \sim M_{\rm dyn}^{\,0.4}$.
\section{Correlation of the B components in the TNG50 sample with $\Riso>1$M\lowercase{pc}}\label{sec:B_corr_with_Riso1_T26}
In \cref{sec:B_v_mass_corr_TNG} we studied correlations of the ordered and turbulent components of the magnetic field and their ratio as well as the total strength for our TNG50 sample of galaxies which also satisfies $\Riso>2$ Mpc. Here, we provide \cref{tab:TNG_vs_T16_Riso1} to list the results for galaxies with $\Riso>1$ Mpc in the TNG50 sample. Excluding the only identified three quenched ones in this sample, gives a total of 490 galaxies. Comparing this table with \cref{tab:TNG_vs_T16}, reveals this point that for galaxies in relatively more involved regions, the aforementioned correlations are weaker.
\section{The rotation curves of galaxies with $\Riso>2$M\lowercase{pc}}\label{sec:rotation_curve}
\begin{figure*}
	\centering
	\includegraphics[scale=0.45]{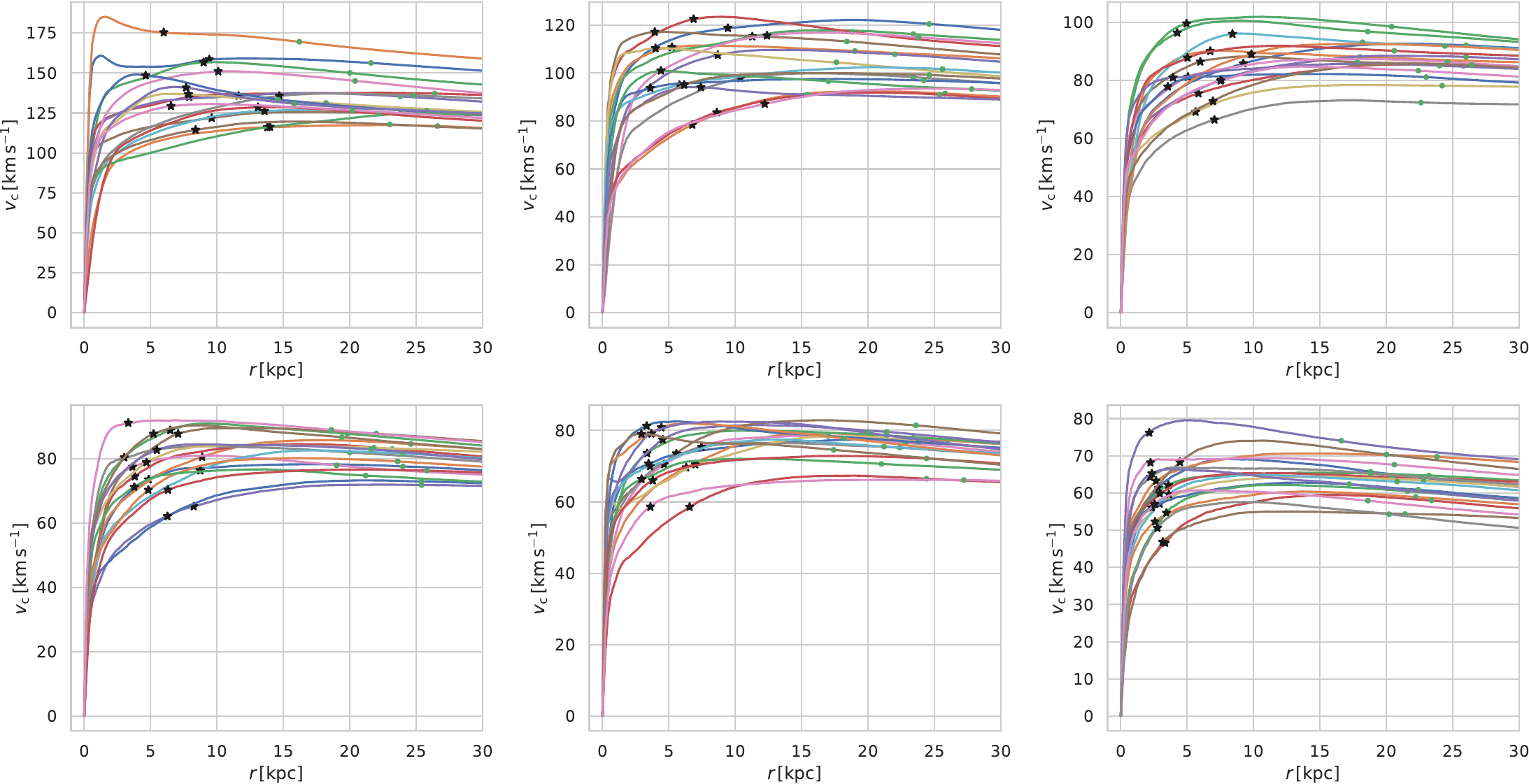}
	\caption{The rotation curves of all the star-forming disc galaxies with $\Riso>2$ Mpc. For each galaxy, $\re$ is denoted by a black star over its rotation curve. The green circles show the points where $v_{\mathrm{c}}=\vrot$ for each galaxy.}
	\label{fig:rcurve}
\end{figure*}

In \cref{sec:compare_with_T16} we calculated $\vrot$ for a sample of 103 galaxies in our TNG50 sample that fulfill $\Riso>2$ Mpc. This was extended to galaxies with $\Riso>1$ Mpc in \cref{sec:B_corr_with_Riso1_T26}. We computed $\vrot$ by taking the average of $v_c$ values (see \cref{eq:vc}) which are calculated at 100 points between the peak of the rotation curve, i.e. where the maximum of $v_\mathrm{c}$ occurred and the $r=30$ kpc distance. In this appendix, we plot \cref{fig:rcurve} to show the rotation curves of all these 103 galaxies. On each curve, the effective radius $\re$ is denoted by a small black star. We also denote the radius at where $v_\mathrm{c}=\vrot$ by a green circle.
\section{Effect of resolution on the $B-M$ relation}\label{sec:effect_of_low_res}
In this work, our findings were based on the highest available run of the TNG simulation suite, i.e. the TNG50-1. To see, how the resolution could affect our results, here, we utilize its lower resolution box TNG50-2, which has the same initial conditions but two and eight times lower spatial and mass resolutions. Like our TNG50-1 sample, we restrict the TNG50-2 galaxies to those of centrals with $\Mstar>10^8$ \msun and then try to find the lower resolution galaxy counterparts of our higher resolution TNG50-1 sample. For this purpose, using the nearest neighbour method, for the star-forming galaxies in our TNG50-1 sample, we search for their first nearest neighbours in the TNG50-2 box and find their corresponding galaxies in this box. We also apply a strict neighbour distance limit of 83 kpc, which yields 1560 equivalent galaxies in these two boxes ($\simeq95\%$ of 1639 galaxies in the TNG50-1). Now, we can compare these two samples.

\begin{figure}
	\centering
	\includegraphics[scale=0.42]{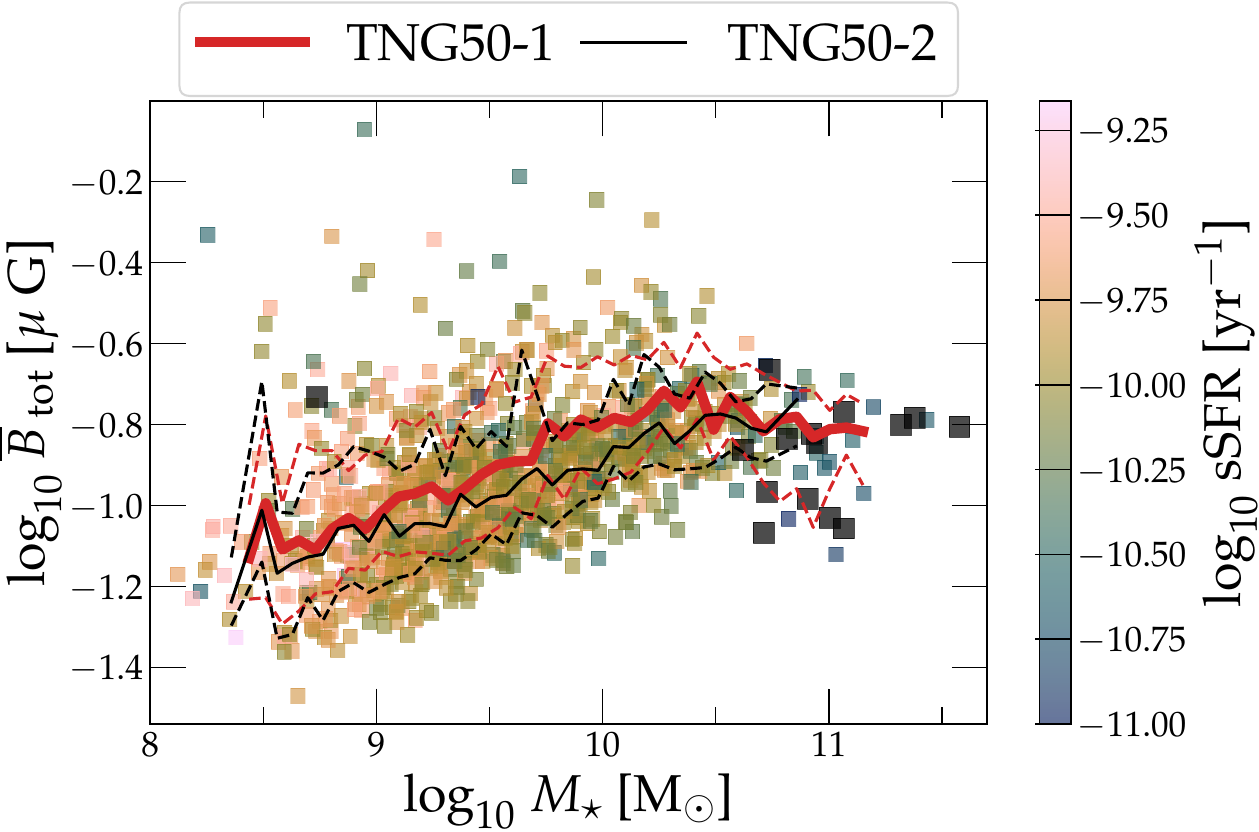}
	\caption{Comparison of $\btot-\Mstar$ relation in TNG50-1 and TNG50-2. For each sample, the solid line shows the median and the upper and lower dashed lines determine the the 84th and 16th percentiles, respectively. Medians for bins with data points $<5$ are not shown. The larger black squares means the galaxy is quenched (see \cref{sec:sf_or_q} for the definition).}
	\label{fig:B_Mstar_ssfr_501_502}
\end{figure}

\cref{fig:B_Mstar_ssfr_501_502} represents an example where we compare $\btot-\Mstar$ relation between the two runs. We can see that in the TNG50-2, the median of $\btot$ is a bit smaller than the TNG50-1, though both show increasing trends. The difference is at most $\sim 0.1$ dex. However, for the more massive galaxies in the TNG50-2, $\btot$ has a decreasing trend. The  difference between the order of the present-day magnetic field in the two runs, could show the importance of resolution on the amplification of the magnetic field in galaxy formation simulations.

%%%%%%%%%%%%%%%%%%%%%%%%%%%%%%%%%%%%%%%%%%%%%%%%%%
% Don't change these lines
\bsp	% typesetting comment
\label{lastpage}
\end{document}